\documentclass[11pt,preprintnumbers,titlepage,nofootinbib]{revtex4-2}
\usepackage[latin9]{inputenc}
\usepackage{amsmath}
\usepackage{amssymb}
\usepackage{graphicx}
\usepackage{esint}
\usepackage[bookmarks=false,
 breaklinks=false,pdfborder={0 0 1},colorlinks=false]
 {hyperref}
\hypersetup{
 colorlinks,linkcolor=blue,citecolor=blue,urlcolor=blue}

\makeatletter

\DeclareTextSymbolDefault{\textquotedbl}{T1}

\usepackage{wasysym}
\usepackage{array}
\usepackage{multirow}
\usepackage{wasysym}
\usepackage{wasysym}
\usepackage{amsfonts}
\usepackage{subfigure}
\usepackage{cleveref}
\usepackage{listings}
\usepackage{xcolor}
\usepackage{array}
\usepackage{url}
\usepackage{color}
\@ifundefined{textcolor}{}{
\definecolor{BLACK}{gray}{0}
\definecolor{WHITE}{gray}{1}
\definecolor{RED}{rgb}{1,0,0}
\definecolor{GREEN}{rgb}{0,1,0}
\definecolor{BLUE}{rgb}{0,0,1}
\definecolor{CYAN}{cmyk}{1,0,0,0}
\definecolor{MAGENTA}{cmyk}{0,1,0,0}
\definecolor{YELLOW}{cmyk}{0,0,1,0}
}

\makeatother

\begin{document}
\title{Hairless Black Hole by Superradiance}

\author{Sebastian Garcia-Saenz$^{a}$}
\email{sgarciasaenz@sustech.edu.cn}

\author{Guangzhou Guo$^{a}$}
\email{guogz@sustech.edu.cn}

\author{Peng Wang$^{b}$}
\email{pengw@scu.edu.cn}

\author{Xinmiao Wang$^{a}$}
\email{12132943@mail.sustech.edu.cn}

\affiliation{$^{a}$Department of Physics, Southern University of Science and Technology, \looseness=-1\\
	Shenzhen 518055, China}

\affiliation{$^{b}$Center for Theoretical Physics, College of Physics, Sichuan University, \looseness=-1\\
	Chengdu 610064, China}

\begin{abstract}
We investigate the interplaying effects of black hole scalarization and superradiance in the context of the Einstein-Maxwell-scalar model, with the scalar field possessing electric charge. Restricted to spherical symmetry, our linear analysis about a Reissner-Nordstr\"om background confirms the persistence of tachyonic scalar modes upon inclusion of electric charge. However, fully nonlinear numerical simulations reveal that the system no longer evolves into a scalarized, hairy black hole state. Instead, we find that the superradiance phenomenon (specifically the electromagnetic version of the effect) causes the scalar condensate to become fully depleted through accretion into the black hole and radiation to spatial infinity. The combination of the two effects, which we refer to as ``tachyonic superradiance'', may thus be seen as a particularly efficient mechanism for the extraction of energy from a black hole, exploiting both the tachyonic growth and superradiant emission. We accurately compute the amounts of energy transfer in different channels by deriving formulae for the energy fluxes of matter fields in the presence of a dynamical horizon, which are amenable for evaluation in numerical relativity calculations.
\end{abstract}

\maketitle

\tableofcontents{}

\section{Introduction}

The amplification of light fields in the vicinity of a black hole (BH) is an interesting scenario in the prospect of probing the existence of new light particles as well as testing their influence on the gravitational field, and hence potentially the dynamics of gravity itself \cite{Arvanitaki:2009fg,Arvanitaki:2010sy,Brito:2014wla,Ferreira:2017pth,Baumann:2018vus,Zhang:2019eid,Garcia-Saenz:2021uyv,Boskovic:2024fga,Brito:2025ojt}. Two well-studied mechanisms leading to the growth of the field amplitude are the scalarization and superradiance effects.

Spontaneous scalarization (see \cite{Doneva:2022ewd} for a review), which as the name indicates applies specifically to scalar fields, concerns the scenario in which a field develops a negative potential well as a result of its coupling with the nontrivial curvature of the BH background, and possibly with other matter fields. The trivial vacuum state of the scalar field is thus subject to a tachyon-type destabilization and the subsequent formation of a condensate, or `scalar cloud'. Naturally this effect is not specific to BHs, and in fact the mechanism was originally described in the setting of neutron stars, i.e.\ the so-called matter-induced scalarization \cite{Damour:1993hw}. It was later understood however that the same effect may be realized in vacuum (i.e.\ without matter fields beyond the scalar) via curvature-induced scalarization, for which scalar-Gauss-Bonnet theory provides a compelling setup \cite{Sotiriou:2013qea,Antoniou:2017hxj,Silva:2017uqg}. Being applicable to vacuum BHs, this description offers a cleaner realization of the mechanism, not to mention that the BHs with scalar hair (i.e.\ the stabilized endpoint state of the scalar cloud) predicted by this model are very interesting systems in their own right.

Superradiance (see \cite{Bekenstein:1998nt,Brito:2015oca} for reviews) refers to the amplification of waves by a dissipative object or medium. Perhaps most familiar and intuitive is the superradiant enhancement of a wave scattering off a rotating object, in which the wave gains energy at the expense of the rotational motion of the object. Although again not peculiar to them, BHs provide a clean and interesting arena to study the effect, as the existence of an ergoregion allows for the extraction of energy (as well as angular momentum and charge) from a BH by a wave scattering off its gravitational potential. Superradiance may also lead to a destabilization process in the event that the magnified scattered waves are continuously reflected back to the BH or are otherwise confined in some way \cite{Press:1972zz,Herdeiro:2013pia,Cardoso:2004hs,Cardoso:2004nk}.

In this paper, we wish to understand the details and consequences of the interplay between the two effects of scalarization and superradiance. As both mechanisms act so as to increase the scalar field amplitude, this combined process, which we refer to as ``tachyonic superradiance'', may be expected to provide an interesting channel for the generation of sizeable radiation out of an initial small perturbation of a (electro-)vacuum BH. On the other hand, as the two effects are associated with different physics, and with different timescales in particular, a careful study of the time evolution and energy transfer among the various fields in the spacetime appears to be warranted. In particular, we discover that the final state of this process is a hairless BH, at least within the model we consider. Tachyonic superradiance may therefore be seen as a particularly efficient mechanism for the extraction of energy from a BH, taking advantage of both the tachyonic growth and superradiant emission.\footnote{A different mechanism to boost superradiant emission has been discussed in \cite{Hui:2022sri,Guo:2025dkx}, which takes advantage of matter accretion. Relatedly, Ref.\ \cite{Cardoso:2013opa} performed a linear analysis of a system that exhibits superradiance and matter-induced scalarization, although not both in conjunction. Finally, we mention the unrelated mechanism of ``descalarization'' \cite{Silva:2020omi,Corelli:2021ikv,Liu:2022eri,Zhang:2022cmu}, which can also result in the depletion of a scalar cloud.}

Our study is restricted to spherical symmetry within the Einstein-Maxwell-scalar (EMS) model \cite{Stefanov:2007eq,Herdeiro:2018wub}. This theory provides a particularly convenient and minimal setup for investigating the combined effects of scalarization and superradiance. On the one hand, the model exhibits tachyonic instabilities similarly to the curvature-induced scalarization models mentioned above, however it achieves so through a nonminimal coupling between the scalar and the electromagnetic field, while all fields are only minimally coupled to standard Einstein gravity. This makes the model theoretically appealing while also allowing one to avoid the complications introduced by higher-order curvature terms. On the other hand, the scalar field in the EMS system is also subject to superradiance if both the BH and the field are charged under the Maxwell $U(1)$ symmetry. This electromagnetic version of the superradiance effect \cite{DiMenza:2014vpa,Benone:2015bst,Baake:2016oku} provides again a cleaner arena for our researches, allowing us to bypass the complications related to BH rotation and the Kerr metric.

We refer to our setup as the `charged EMS' model, to emphasize the fact that the scalar field is electrically charged, requiring therefore the extension of the original proposal to the case where the scalar field is complex. This also provides a natural generalization of the setup studied recently in \cite{Garcia-Saenz:2024beb}, where the field was assumed complex but neutral. Naturally, in this model, both mechanisms of scalarization and superradiance necessitate the existence of BH electric charge, which is not expected to be sizeable in the astrophysical context. As an alternative scenario, the gauge field of the EMS system need not correspond to the Standard Model photon but may instead be associated with a dark $U(1)$ sector \cite{Essig:2013lka}. One can speculate that astrophysical BHs might acquire a charge corresponding to this $U(1)$ symmetry and thus support the scalarization and superradiance effects that we investigate here.

It is worth commenting on the connection and differences between the asymptotically flat setup that we focus on and the case of asymptotically anti-de Sitter (AdS) BHs within the EMS model, which has been extensively studied in the AdS/CFT literature, particularly in the context of holographic superconductors (see \cite{Hartnoll:2009sz,Herzog:2009xv,Horowitz:2010gk} for reviews). Although both setups exhibit superradiant amplification of the scalar field, only the asymptotically AdS model features an effective potential barrier that allows for bound states and thus the formation of a scalar condensate. Moreover, the existence of bound states requires a minimum depth of the potential well, leading to a non-trivial dependence of the scalar amplitude on the temperature of the BH, in particular the existence of a critical temperature above which no condensation occurs. In contrast, the asymptotically flat case is characterized by the absence of a potential barrier, and hence the lack of a confining mechanism for superradiantly amplified waves \cite{Hod:2013nn,Hod:2015hza}. Nevertheless, one may still expect a nontrivial dependence of observables on the temperature of the system (equivalently, in the flat case, on the charge-to-mass ratio of the BH), as we will see to be the case for the amounts of energy and charge loss resulting from the tachyonic superradiance mechanism.

Our setup is described in detail in Sec.\ \ref{secII}. We begin with a description of the model and the derivation of the equations of motion in the $3+1$ decomposition that we employ in our numerical relativity calculations. As we are especially interested in the details of energy and charge transfer toward spatial infinity and between the different degrees of freedom in the system, we provide explicit derivations of expressions for the total energy and charge of the BH as well as their fluxes, which are well-suited for evaluation in numerical simulations. In particular, we generalize existing formulae for the local continuity of energy to the situation where the BH horizon is expanding. Finally, we derive the linearized equation for scalar modes and discuss the boundary conditions relevant for determining the presence of tachyonic and superradiant effects. Sec.\ \ref{sec:Stable-Evolution-in} presents our numerical results and their analysis. As our main objective is to highlight the characteristic properties of the tachyonic superradiance mechanism, we start for the sake of comparison with the study of some simpler systems which exhibit either superradiance or scalarization, but not both, before turning to the general case. Some final comments and prospects are offered in Sec.\ \ref{Sec:Conc}. The Appendices contain some technical information related to our numerical calculations.

%%%%%%%%%%%%%%%%%%%%%%%%%%%%%%
%%%%%%%%%%%%%%%%%%%%%%%%%%%%%%

\section{Setup}
\label{secII}

We describe in this section the details of the charged EMS model and the fully nonlinear evolution equations to be solved in our numerical simulations. We provide approximate expressions for the BH mass and charge valid in spherically symmetric but time-dependent systems, which are suitable for numerical evaluation. We next derive a continuity equation for curvature and matter energies in the spacetime region between a dynamical horizon and a distant observer, generalizing previous work to include the case when the spacetime possesses no time-like Killing vector. Finally, we derive the linearized equation and appropriate boundary conditions from which the existence of tachyonic instabilities and superradiance of the scalar field may be studied.

\subsection{Charged EMS model}

The EMS model consists of a scalar field $\Phi$ minimally coupled to Einstein gravity and nonminimally coupled to the electromagnetic field $A_\mu$. In the original setup studied in the context of BH scalarization \cite{Herdeiro:2018wub,Fernandes:2019rez,Minamitsuji:2021vdb,Guo:2023mda,Melis:2024kfr}, the scalar field is real and uncharged under the electromagnetic $U(1)$ gauge symmetry. Here we consider the inclusion of electric charge for the scalar, i.e.\ $\Phi$ is complex in our setup. The EMS model with complex scalar field, but without charge, was first investigated in \cite{Garcia-Saenz:2024beb}.\footnote{The EMS model with an additional, spectator-like complex scalar was studied in \cite{Corelli:2021ikv}. See also \cite{Latosh:2023cxm,Hyun:2024sfv} in the context of scalar-Gauss-Bonnet theory.} The action is given by\footnote{We use geometrized units with $G=c=4\pi\epsilon_0=1$. Note that the matter fields are not canonically normalized.}
\begin{equation}
S_{\rm EMS}=\frac{1}{16\pi}\int d^{4}x\sqrt{-g}\left[R-2\left(\mathcal{D}_{\mu}\Phi\right)^{*}\mathcal{D}^{\mu}\Phi-f\left(|\Phi|\right)F^{\mu\nu}F_{\mu\nu}\right] \,, \label{eq:action}
\end{equation}
in terms of $F_{\mu\nu}\equiv \partial_{\mu}A_{\nu}-\partial_{\nu}A_{\mu}$ and
$\mathcal{D}_{\mu}\equiv\nabla_{\mu}-iqA_{\mu}$, where $q$ is the charge of the scalar field. We adopt the choice $f\left(|\Phi|\right)=e^{\alpha_{0}|\Phi|^2}$, where $\alpha_0$ is a real constant, for the nonminimal coupling function in \eqref{eq:action}, which allows for hairless RN BH solutions.

Varying the action \eqref{eq:action} with respect to the scalar and electromagnetic fields yields the following equations of motion (here $F^2\equiv F^{\mu\nu}F_{\mu\nu}$):
\begin{align}
\mathcal{D}^{\mu}\mathcal{D}_{\mu}\Phi & =\frac{1}{2}\alpha_{0}e^{\alpha_{0}|\Phi|^2}F^{2}\Phi \,,\label{eq:scalar_eq}\\
\nabla_{\nu}F^{\mu\nu} & =4\pi j_{e}^{\mu} \,,\label{eq:A_eq}
\end{align}
in terms of the electric current density
\begin{equation}
j_{e}^{\mu}=-\frac{e^{-\alpha_{0}|\Phi|^2}}{4\pi}\left[F^{\mu\nu}\partial_{\nu}\left(e^{\alpha_{0}|\Phi|^2}\right)+\frac{1}{2}iq\left(\Phi^{*}\partial^{\mu}\Phi-\Phi\partial^{\mu}\Phi^{*}\right)+q^{2}A^{\mu}|\Phi|^2\right] \,.
\end{equation}

To the end of investigating the fully nonlinear dynamics of the EMS model, we adopt a $3+1$ decomposition of spacetime in terms of the ADM metric
\begin{equation}
ds^{2}=-\alpha^{2}dt^{2}+\gamma_{ij}\left(dx^{i}+\beta^{i}dt\right)\left(dx^{j}+\beta^{j}dt\right) \,,\label{metric}
\end{equation}
where $\alpha$ is the lapse function, $\beta^{i}$ the shift vector and $\gamma_{ij}$ the induced physical metric on the 3-dimensional spatial slice $\Sigma$ \cite{Baumgarte:2010ndz,Frauendiener:2011zz}. The normal vector to $\Sigma$ is $n_{\mu}=\left(-\alpha,0,0,0\right)$, or $n^{\mu}=\left(1/\alpha,-\beta^{i}/\alpha\right)$, and $\gamma_{\mu}^{\nu}=\delta_{\mu}^{\nu}+n_{\mu}n^{\nu}$ defines a projector which maps a 4-vector onto $\Sigma$.

The matter variables are similarly decomposed. We introduce the variable $\Pi=n^{\mu}\nabla_{\mu}\Phi$ related to the momentum of the scalar field, as well as the projections 
\begin{equation}
\mathcal{A}_{*}=-n^{\mu}A_{\mu}\,,\quad\mathcal{A}_{i}=\gamma_{i}^{\mu}A_{\mu}\,,
\end{equation}
of the Maxwell field. The electric and magnetic fields are also defined in terms of projections of $F_{\mu\nu}$,
\begin{equation}
E_{i}=\gamma_{i}^{\mu}n^{\nu}F_{\mu\nu}\,,\quad B_{i}=\gamma_{i}^{\mu}n^{\nu}\widetilde{F}_{\mu\nu}\,,
\end{equation}
where $\widetilde{F}_{\mu\nu}$ is the Hodge dual of $F_{\mu\nu}$. With these definitions, and adopting the Lorenz gauge $\nabla_{\mu}A^{\mu}=0$, the matter equations \eqref{eq:scalar_eq} and \eqref{eq:A_eq} may be recast into the following set of equations:
\begin{align}
\left(\partial_{t}-\mathcal{L}_{\beta}\right)\Phi &= \alpha\Pi \,,\nonumber \\
\left(\partial_{t}-\mathcal{L}_{\beta}\right)\Pi &= D_{i}\left(\alpha\partial^{i}\Phi\right)+\alpha\Pi K-2iq\alpha\left(\mathcal{A}_{i}\partial^{i}\Phi+\mathcal{A}_{*}\Pi\right)\nonumber \\
 &\quad -\alpha\left[\frac{1}{2}\alpha_{0}e^{\alpha_{0}|\Phi|^2}F^{2}\Phi+q^{2}\left(\mathcal{A}_{i}\mathcal{A}^{i}-\mathcal{A}_{*}^{2}\right)\Phi\right] \,,\label{eq:matter_eqs}\\
\left(\partial_{t}-\mathcal{L}_{\beta}\right)E^{i} &= \alpha KE^{i}+\alpha\epsilon^{ijk}D_{j}B_{k}-\epsilon^{ijk}B_{j}D_{k}\alpha-4\pi\alpha{}^{(3)}j_{e}^{i} \,,\nonumber \\
\left(\partial_{t}-\mathcal{L}_{\beta}\right)\mathcal{A}_{i} &= -\alpha\left(E_{i}+D_{i}\mathcal{A}_{*}\right)-\mathcal{A}_{*}D_{i}\alpha \,,\nonumber \\
\left(\partial_{t}-\mathcal{L}_{\beta}\right)\mathcal{A}_{*} &= -\mathcal{A}^{i}D_{i}\alpha+\alpha\left(K\mathcal{A}_{*}-D_{i}\mathcal{A}^{i}\right) \,,\nonumber 
\end{align}
where $\mathcal{L}_{\beta}$ is the Lie derivative along the shift vector $\beta^{i}$, $D_i$ and $\epsilon_{ijk}$ are respectively the covariant derivative and Levi-Civita tensor associated with $\gamma_{ij}$, and $K=\gamma^{ij}K_{ij}$ is the trace of the extrinsic curvature on $\Sigma$ \cite{Torres:2014fga,Sanchis-Gual:2016tcm,Hirschmann:2017psw}. The electric charge density $\rho_{e}$ and spatial current density $^{(3)}j_{e}^{i}$ are given by
\begin{equation}
\rho_{e}=-n_{\mu}j_{e}^{\mu} \,,\quad{}^{(3)}j_{e}^{i}=\gamma_{\mu}^{i}j_{e}^{\mu} \,.
\end{equation}

Concerning the numerical evolution of the Einstein equations, we utilize the so-called CCZ3 formulation, a recently developed numerical scheme that has been shown to provide robust and accurate simulations of BH spacetimes; see \cite{Garcia-Saenz:2025dsr} for the relevant equations and further explanations. The equations require as input the $3+1$ components of the stress-energy tensor, defined as
\begin{equation}
\rho=n^{\mu}n^{\nu}T_{\mu\nu} \,,\quad S_{i}=-\gamma_{i}^{\mu}n^{\nu}T_{\mu\nu} \,,\quad S_{ij}=\gamma_{i}^{\mu}\gamma_{j}^{\nu}T_{\mu\nu} \,,
\end{equation}
where in the charged EMS model one has
\begin{equation}
T_{\mu\nu}=\frac{1}{8\pi}\left[\left(\mathcal{D}_{\mu}\Phi\right)^{*}\mathcal{D}_{\nu}\Phi+\left(\mathcal{D}_{\nu}\Phi\right)^{*}\mathcal{D}_{\mu}\Phi-g_{\mu\nu}\left(\mathcal{D}^{\rho}\Phi\right)^{*}\mathcal{D}_{\rho}\Phi+e^{\alpha_{0}|\Phi|^2}\left(2F_{\mu\rho}F_{\nu}{}^{\rho}-\frac{1}{2}g_{\mu\nu}F^{2}\right)\right]\,.
\end{equation}

%%%%%%%%%%%%%%%%%%%%%%%%%%%%%%%%%%%%%%%

\subsection{Black hole mass and charge}

In asymptotically flat spacetimes, the ADM mass quantifies the total energy of the spacetime, including contributions from both curvature and matter \cite{Misner:1974qy}. On a constant-time slice $\Sigma_{t}$, the ADM mass may be defined as \cite{Poisson:2009pwt}
\begin{equation}
M_{\rm ADM}=-\frac{1}{8\pi}\lim_{S_{t}\rightarrow\infty}\oint_{S_{t}}d^{2}\theta\sqrt{\sigma}\left(k-k_{0}\right) \,, \label{eq:M_ADM}
\end{equation}
where $S_{t}$ is a two-sphere at spatial infinity (as implied by the limit), $k$ is the trace of the extrinsic curvature of $S_{t}$ embedded in $\Sigma_{t}$, and $k_{0}$ is the trace of the extrinsic curvature of $S_{t}$ embedded in flat spacetime. 

In this paper, we focus on spherically symmetric spacetimes, so that the line element $dl^{2}$ on $\Sigma_{t}$ can be expressed in terms of the areal radius $R$ as
\begin{equation}
dl^{2}=\frac{1}{1-\frac{2M\left(R\right)}{R}}dR^{2}+R^{2}d\Omega^{2} \,, \label{eq:dl2}
\end{equation}
so that the ADM mass can be expressed as
\begin{equation}
M_{\rm ADM}=\lim_{R\rightarrow\infty}M\left(R\right) \,.
\end{equation}
Since the energy density of matter fields decays as a power law at spatial infinity, evaluating $M\left(R\right)$ at a sufficiently large $R$ provides a convenient measure for the total energy of the spacetime, which we denote by $M_{\rm BH}$, in practical numerical simulations \cite{Torres:2014fga}.

To determine the BH charge in a time-dependent system, we first consider the continuity equation that follows from \eqref{eq:A_eq},
\begin{equation}
\nabla_{\mu}j_{e}^{\mu}=0 \,. \label{eq:A_contEq}
\end{equation}
In terms of the $3+1$ decomposition variables, the left hand side of \eqref{eq:A_contEq} reads
\begin{equation}
\nabla_{\mu}j_{e}^{\mu}=\alpha^{-1}\left[D_{i}\left(\alpha{}^{(3)}j_{e}^{i}-\rho_{e}\beta^{i}\right)+\partial_{t}\rho_{e}\right] \,,
\end{equation}
so that the continuity equation can be recast as
\begin{equation}
\partial_{t}\rho_{e}=D_{i}\left(\rho_{e}\beta^{i}-\alpha{}^{(3)}j_{e}^{i}\right) \,.\label{eq:cont_drhoE}
\end{equation}
In spherically symmetric spacetimes, the electric charge $Q\left(r\right)$ enclosed within a 2-sphere $\Sigma\left(r\right)$ of radius $r$ (in some suitable coordinates) is defined as 
\begin{equation}
Q\left(r\right)=\int_{\Sigma(r)}dV\,\rho_{e} \,. \label{eq:def_Q}
\end{equation}
Considering the time derivative of this expression and employing \eqref{eq:cont_drhoE} along with Gauss's law, we obtain \cite{Torres:2014fga}
\begin{equation}
\frac{dQ\left(r\right)}{dt}=\int_{\partial\Sigma(r)}dS\,\hat{r}_{i}\left(\rho_{e}\beta^{i}-\alpha{}^{(3)}j_{e}^{i}\right)=4\pi\sqrt{\gamma_{rr}}\gamma_{\theta\theta}\left(\rho_{e}\beta^{r}-\alpha{}^{(3)}j_{e}^{r}\right) \,,
\end{equation}
where we make use of the spherical symmetry and the corresponding form of the unit normal vector $\hat{r}_{i}=\left(\gamma_{rr},0,0\right)$ to arrive at the last equality. 

Summarizing, the mass and charge measured by a distant observer at $r=R_{obs}$ in a time-dependent BH spacetime
can be approximated as
\begin{align}
M_{\rm BH}\left(t\right) & =\left.M\left(t\right)\right|_{r=R_{obs}} \,,\label{eq:BHmass}\\
Q_{\rm BH}\left(t\right) & =Q_{0}+\left.4\pi\int_{t_{0}}^{t}dt\sqrt{\gamma_{rr}}\gamma_{\theta\theta}\left(\rho_{e}\beta^{r}-\alpha{}^{(3)}j_{e}^{r}\right)\right|_{r=R_{obs}} \,,\label{eq:BHcharge}
\end{align}
where $M\left(t\right)$ is evaluated using Eq.\ \eqref{eq:dl2}, and $Q_{0}$ is the initial BH charge. In our numerical simulations, the observer's location will coincide with the spatial boundary of our numerical domain, and the changes in $M_{\rm BH}$ and $Q_{\rm BH}$ may be used to measure the energy and charge radiated to infinity by the fields.

%%%%%%%%%%%%%%%%%%%%%%%%%%%%%%%%%%%%%%%%

\subsection{Local continuity of energy}

\begin{figure}[t]
\begin{centering}
\includegraphics[scale=0.8]{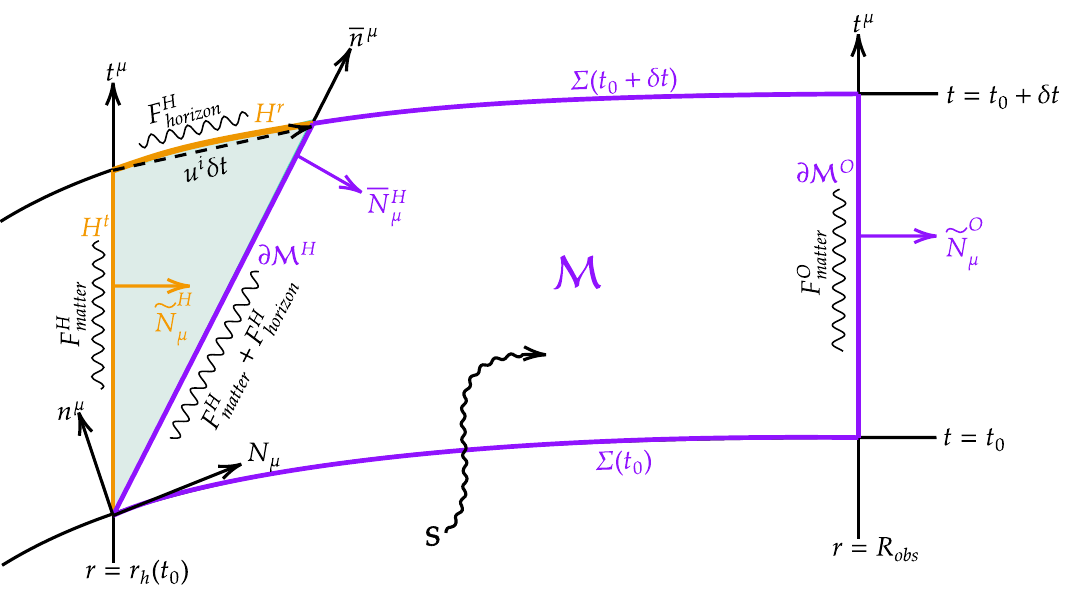}
\par\end{centering}
\caption{Representation of energy flows during BH evolution. The 4-dimensional manifold $\mathcal{M}$ represents the spacetime exterior to the BH in a finite-domain simulation, which is bounded by the purple lines: the time-dependent horizon surface $\partial\mathcal{M}^{H}$, the hypersurface of a distant
observer $\partial\mathcal{M}^{O}$, and two spatial slices $\Sigma\left(t_{0}\right)$ and $\Sigma\left(t_{0}+\delta t\right)$. The energy flux through the BH horizon surface turns out to be the combination of the matter flux $F_{\rm matter}^{H}$ through the horizon and the flux $F_{\rm horizon}^{H}$ due to the dynamical expansion of the horizon. Additionally, an extra energy source $S$ arises from the conversion between matter and curvature energies, which occurs due to the absence of a time-like Killing vector in a dynamical system. Further details on the diagram are given in the main text.}
\label{contEnergy}
\end{figure}

In analogy with the electromagnetic continuity equation \eqref{eq:cont_drhoE}, we aim to derive the corresponding continuity equation for matter and curvature energy, relating the energy current and flux within a local region. We begin by defining a current $J^{\mu}$ by projecting the stress-energy tensor $T_{\mu\nu}$ onto the time-like direction
$t^{\mu}=\left(1,0,0,0\right)$,
\begin{equation}
J^{\mu}=t^{\nu}T^{\mu}{}_{\nu} \,.\label{eq:def_J}
\end{equation}
Energy-momentum conservation implies $\nabla_{\mu}T^{\mu\nu}=0$, however it does not follow that $\nabla_{\mu}J^{\mu}=0$ because $t^{\mu}$ is not a Killing vector in a dynamical spacetime. Integrating the divergence of the previous expression we have
\begin{equation}
\int_{\mathcal{M}}d^{4}x\sqrt{-g}\,\nabla_{\mu}J^{\mu}=-\int_{t_{0}}^{t}dt\int_{\Sigma\left(t\right)}d^{3}x\sqrt{\gamma}\,\mathcal{S} \,,\label{eq:4-contEq}
\end{equation}
where $\mathcal{M}$ is the 4-dimensional manifold bounded by spatial hypersurfaces $\Sigma(t_0)$ and $\Sigma(t)$, and $\mathcal{S}$ represents the additional source density, given by
\begin{equation}
\mathcal{S}=-\alpha T_{\mu\nu}\nabla^{\mu}t^{\nu}=\rho\partial_{t}\alpha-S_{i}\partial_{t}\beta^{i}-\frac{\alpha}{2}S^{ij}\partial_{t}\gamma_{ij} \,.
\end{equation}
The physical implication of $\mathcal{S}$, as described in \cite{Clough:2021qlv,Croft:2022gks}, is that it characterizes the energy transfer from the curvature to the matter sector. 

The dynamical BH setup is represented in Fig.\ \ref{contEnergy}. The manifold $\mathcal{M}$ is bounded by two hypersurfaces $\partial\mathcal{M}^{H}$ (the dynamical BH horizon) and $\partial\mathcal{M}^{O}$ (the hypersurface of a distant observer), along with the two 3-spatial slices $\Sigma\left(t_{0}\right)$ and $\Sigma\left(t\right)$ already mentioned; we write $t=t_0+\delta t$ and we will eventually take the limit $\delta t\to0$. For latter reference, we denote the 2-dimensional spatial surface corresponding to the BH horizon by $\partial H$, which is located at $r=r_{h}(t)$, and the 2-sphere for a distant observer at $r=R_{obs}$ by $\partial\Sigma^{O}$. Note that $\partial H$ is not necessarily spherical in this discussion (although it will be in our calculations). We shall also need to be careful in distinguishing different sets of coordinates with specific notations:
\begin{itemize}
\item $x^{\mu}=\left(t,r,\theta,\varphi\right)$ for the 4-dimensional manifold $\mathcal{M}$; 
\item $x^{i}=\text{\ensuremath{\left(r,\theta,\varphi\right)}}$ for the 3-dimensional spatial hypersurfaces ($\Sigma\left(t_{0}\right)$, $\Sigma\left(t_{0}+\delta t\right)$ and $H^{r}$);
\item $X^{a}=\text{\ensuremath{\left(t,\theta,\varphi\right)}}$ for the 3-dimensional time-like hypersurfaces ($\partial\mathcal{M}^{H}$,
$\partial\mathcal{M}^{O}$ and $H^{t}$); 
\item $\theta^{m}=\left(\theta,\varphi\right)$ for the 2-dimensional surfaces ($\partial H$ and $\partial\Sigma^{O}$). 
\end{itemize}
Lastly we provide a detailed description of the hypersurfaces depicted in Fig.\ \ref{contEnergy}, along with some additional definitions:
\begin{itemize}
\item $\Sigma\left(t_{0}\right)$: This is the 3-dimensional spatial slice normal to the unit vector $n^{\mu}$. The unit vector $N^{\mu}=\left(0,N^{i}\right)$
is normal to $n^{\mu}$, i.e.\ it satisfies $n_{\mu}N^{\mu}=0$ and $g_{\mu\nu}N^{\mu}N^{\nu}=\gamma_{ij}N^{i}N^{j}=1$. The induced metric on $\Sigma\left(t_{0}\right)$ is given by $\gamma_{ij}$.

\item $\Sigma\left(t_{0}+\delta t\right)$: This is the spatial slice obtained by evolving $\Sigma\left(t_0\right)$ along the normal vector $n^{\mu}$ through a time step $\delta t$. The normal vector and the induced metric are $n^{\mu}$ and $\gamma_{ij}$ at $t=t_{0}+\delta t$, respectively.

\item $\partial\mathcal{M}^{H}$: This is the $2+1$ hypersurface representing the evolution of the BH horizon $\partial H$ along the direction
$\bar{n}^{\mu}\propto\left(1,u^{i}\right)$, where $u^{i}$ is the expansion velocity of $\partial H$. The unit vector $\bar{N}_{\mu}^{H}=\left(-u^{i}\bar{N}_{i}^{H},\bar{N}_{i}^{H}\right)$ is normal to $\bar{n}^{\mu}$, and the normalization condition $g^{\mu\nu}\bar{N}_{\mu}^{H}\bar{N}_{\nu}^{H}=1$ implies
\begin{equation}
\bar{N}_{i}^{H}=\frac{\alpha}{\sqrt{\alpha^{2}-\left(\beta^{i}+u^{i}\right)\left(\beta^{j}+u^{j}\right)N_{i}^{H}N_{j}^{H}}}N_{i}^{H} \,,
\end{equation}
where $N_{\mu}^{H}=N_{\mu}|_{r=r_{h}}$. On the BH surface
$\partial H$, we specify $N_{i}^{H}$ to be its normal vector, so that $\bar{N}_{\mu}^{H}$ is normal to the hypersurface $\partial\mathcal{M}^{H}$. The induced metric $\bar{h}_{\mu\nu}^{H}$ is given by projecting
$g_{\mu\nu}$ onto $\partial\mathcal{M}^{H}$ via $\bar{h}_{\mu\nu}^{H}=g_{\mu\nu}-\bar{N}_{\mu}^{H}\bar{N}_{\nu}^{H}$. The determinant of $\bar{h}_{ab}^{H}=\frac{\partial x^{\mu}}{\partial X^{a}}\frac{\partial x^{\nu}}{\partial X^{b}}\bar{h}_{\mu\nu}^{H}$ can be expressed as 
\begin{equation}
\bar{h}^{H}=-\left[\alpha^{2}-\left(\beta^{i}+u^{i}\right)\left(\beta^{j}+u^{j}\right)N_{i}^{H}N_{j}^{H}\right]\sigma^{H} \,,
\end{equation}
where $\sigma^{H}$ is the determinant of the induced metric $\sigma_{mn}^{H}$ on $\partial H$.

\item $\partial\mathcal{M}^{O}$: This is the $2+1$ hypersurface representing a distant observer at $r=R_{obs}$ evolving along the direction $t^{\mu}=\left(1,0,0,0\right)$. Its normal vector is $\tilde{N}_{\mu}^{O}=\left(0,\tilde{N}_{i}^{O}\right)$ with 
\begin{equation}
\tilde{N}_{i}^{O}=\frac{\alpha}{\sqrt{\alpha^{2}-\beta^{i}\beta^{j}N_{i}^{O}N_{j}^{O}}}N_{i}^{O},\label{eq:N_tilde}
\end{equation}
where $N_{\mu}^{O}=N_{\mu}|_{r=R_{obs}}$, and the direction of $N_{i}^{O}$ is normal to $\partial\Sigma^{O}$. The induced metric $\tilde{h}_{ab}^{O}$ on $\partial\mathcal{M}^{H}$ can be expressed as $\tilde{h}_{\mu\nu}^{O}=g_{\mu\nu}-\tilde{N}_{\mu}^{O}\tilde{N}_{\nu}^{O}$,
and its determinant
\begin{equation}
\tilde{h}^{O}=-\left(\alpha^{2}-\beta^{i}\beta^{j}N_{i}^{O}N_{j}^{O}\right)\sigma^{O},\label{eq:h_tilde}
\end{equation}
where $\sigma^{O}$ is the determinant of the induced metric $\sigma_{mn}^{O}$ on $\partial\Sigma^{O}$.

\item $H^{t}$: This is the $2+1$ hypersurface of $r=r_{h}\left(t_{0}\right)$ representing the evolution of the horizon $\partial H$ along the direction $t^{\mu}$. Its normal vector is $\tilde{N}_{\mu}^{H}=\left(0,\tilde{N}_{i}^{H}\right)$,
where $\tilde{N}_{i}^{H}$ satisfies the relation $\tilde{N}_{i}^{H}=\alpha N_{i}^{H}/\sqrt{\alpha^{2}-\beta^{i}\beta^{j}N_{i}^{H}N_{j}^{H}}$. The induced metric projection $\tilde{h}_{ab}^{H}$ is obtained from
$\tilde{h}_{\mu\nu}^{H}=g_{\mu\nu}-\tilde{N}_{\mu}^{H}\tilde{N}_{\nu}^{H}$,
and has the determinant $\tilde{h}^{H}=-\left(\alpha^{2}-\beta^{i}\beta^{j}N_{i}^{H}N_{j}^{H}\right)\sigma^{H}$.

\item $H^{r}$: This is the 3-dimensional spatial slice representing the motion of the horizon $\partial H$ along the direction $u^{i}$, which has the normal vector $n^{\mu}$ and the induced metric $\gamma_{ij}$ at $t=t_{0}+\delta t$. This slice $H^{r}$ is bounded by the horizon surfaces $\partial H\left(t_{0}\right)$ and $\partial H\left(t_{0}+\delta t\right)$ with a perpendicular interval length $\delta s=N_{i}^{H}u^{i}\delta t$ in the limit $\delta t\to0$, where $N_{i}^{H}$ is normal to $\partial H\left(t_{0}\right)$. 
\end{itemize}

Applying Gauss's law to the left-hand side of \eqref{eq:4-contEq} and making use of the previous definitions, we have
\begin{align}
\int_{\mathcal{M}}d^{4}x\sqrt{-g}\,\nabla_{\mu}J^{\mu}= & \int_{\Sigma\left(t_{0}\right)}d^{3}x\sqrt{\gamma}\left(n_{\mu}J^{\mu}\right)+\int_{\Sigma\left(t_{0}+\delta t\right)}d^{3}x\sqrt{\gamma}\left(-n_{\mu}J^{\mu}\right)\nonumber \\
 & +\int_{\partial\mathcal{M}^{H}}d^{3}X\sqrt{-\bar{h}^{H}}\left(-\bar{N}_{\mu}^{H}J^{\mu}\right)+\int_{\partial\mathcal{M}^{O}}d^{3}X\sqrt{-\tilde{h}^{O}}\left(\tilde{N}_{\mu}^{O}J^{\mu}\right) \,.\label{eq:integrations_partialM}
\end{align}
So far everything is exact. Now we consider the limit $\delta t\to0$. We divide \eqref{eq:4-contEq} by $\delta t$ and consider the different terms on the right-hand side of \eqref{eq:integrations_partialM}, arriving at the following equation expressing the continuity of energy:
\begin{equation}
\frac{dE_{\rm matter}}{dt}=F_{\rm matter}^{H}+F_{\rm horizon}^{H}+F_{\rm matter}^{O}+S \,,\label{eq:energy_contEq}
\end{equation}
where one has the following definitions: 
\begin{itemize}
\item $E_{\rm matter}$ is the total matter energy in the region $\Sigma$ (and note that in the limit $\delta t\to0$ one need not specify the time value of $\Sigma(t)$), which arises from the first two terms on the right-hand side of \eqref{eq:integrations_partialM}, and is given by
\begin{equation}
E_{\rm matter}=\int_{\Sigma}d^{3}x\sqrt{\gamma}\left(n_{\mu}J^{\mu}\right)=\int_{\Sigma}d^{3}x\sqrt{\gamma}\,\mathcal{E}_{matter} \,,
\end{equation}
where $\mathcal{E}_{\rm matter}\equiv \alpha\rho-\beta_{k}S^{k}$ is the energy density of matter.

\item $F_{\rm matter}^{H}$ is the flux of matter energy through the BH horizon 2-surface $\partial H$, and is given by 
\begin{equation}
F_{\rm matter}^{H}=\int_{\partial H}d^{2}\theta\sqrt{\sigma}\left(-\alpha N_{i}^{H}J^{i}\right)=\int_{\partial H}d^{2}\theta\sqrt{\sigma^{H}}\mathcal{F}_{\rm matter}^{H} \,,
\end{equation}
in terms of $\mathcal{F}_{\rm matter}^{H}\equiv -N_{i}^{H}\left[\left(\beta^{i}\left(\alpha\rho-\beta^{j}S_{j}\right)+\alpha\left(\beta^{k}S_{k}^{i}-\alpha S^{i}\right)\right)\right]$.

\item $F_{\rm horizon}^{H}$ is the horizon flux related to the expansion of the horizon itself, and it is expressed as
\begin{equation}
F_{\rm horizon}^{H}=\int_{\partial H}d^{2}\theta\sqrt{\sigma}\left(\alpha u^{i}N_{i}^{H}J^{0}\right)=\int_{\partial H}d^{2}\theta\sqrt{\sigma^{H}}\mathcal{F}_{\rm horizon}^{H} \,,\label{eq:horizon_flux}
\end{equation}
where $\mathcal{F}_{\rm horizon}^{H}\equiv -N_{i}^{H}u^{i}\mathcal{E}_{\rm matter}$. Notice that $F_{\rm matter}^{H}$ and $F_{\rm horizon}^{H}$ come from expanding the third term on the right-hand side of \eqref{eq:integrations_partialM}.

\item $F_{\rm matter}^{O}$ is the flux of matter energy measured by the distant observer at $r=R_{obs}$. It is written as
\begin{equation}
F_{\rm matter}^{O}=\int_{\partial\Sigma^{O}}d^{2}\theta\sqrt{\sigma}\left(\alpha N_{i}^{O}J^{i}\right)=\int_{\partial\Sigma^{O}}d^{2}\theta\sqrt{\sigma^{O}}\mathcal{F}_{\rm matter}^{O},
\end{equation}
with $\mathcal{F}_{\rm matter}^{O}\equiv N_{i}^{O}\left[\left(\beta^{i}\left(\alpha\rho-\beta^{j}S_{j}\right)+\alpha\left(\beta^{k}S_{k}^{i}-\alpha S^{i}\right)\right)\right]$.

\item $S$ represents the integrated source term, which as explained accounts for the energy transfer from the curvature to the matter sector,
\begin{equation}
S=\int_{\Sigma}d^{3}x\sqrt{\gamma}\,\mathcal{S} \,.
\end{equation}
\end{itemize}
Note that the minus sign in the definition of $\mathcal{F}_{\rm matter}^{H}$ means that
$\mathcal{F}_{\rm matter}^{H}<0$ signals an ingoing energy flux through the BH horizon, while $\mathcal{F}_{\rm matter}^{O}<0$ indicates an outgoing energy flux measured by the distant observer.

Additional insight may be gained by performing the surface integration over $H^{t}$ and $H^{r}$ in the limit $\delta t\rightarrow0$,
\begin{align}
\int_{H^{t}}d^{3}X\sqrt{-\tilde{h}^{H}}\left(\tilde{N}_{\mu}^{H}J^{\mu}\right) & =\delta t\int_{\partial H}d^{2}\theta\sqrt{\sigma^{H}}\alpha N_{i}^{H}J^{i}=-\delta t\,F_{\rm matter}^{H} \,,\\
\int_{H^{r}}d^{3}x\sqrt{\gamma}\left(-n_{\mu}J^{\mu}\right) & =-\delta t\int_{\partial H}d^{2}\theta\sqrt{\sigma^{H}}N_{i}^{H}u^{i}\mathcal{E}_{\rm matter}=-\delta t\,F_{\rm horizon}^{H} \,,
\end{align}
which lead to the interesting identity 
\begin{equation}
\lim_{\delta t\rightarrow0}\int_{\partial\mathcal{M}^{H}}d^{3}X\sqrt{-\bar{h}^{H}}\left(-\bar{N}_{\mu}^{H}J^{\mu}\right)=\lim_{\delta t\rightarrow0}\left[\int_{H^{t}}d^{3}X\sqrt{-\tilde{h}^{H}}\left(\tilde{N}_{\mu}^{H}J^{\mu}\right)+\int_{H^{r}}d^{3}x\sqrt{\gamma}\left(-n_{\mu}J^{\mu}\right)\right] \,.
\end{equation}
which is nothing but the statement that the total energy flux through the dynamical BH horizon is the sum of two contributions: the matter flux $F_{\rm matter}^{H}$
directly through the horizon and the horizon flux $F_{\rm horizon}^{H}$ induced by the expansion.

%%%%%%%%%%%%%%%%%%%%%%%%%%%%%%%%%%%%%%%

\subsection{Linear analysis}
\label{subsec:Linear-analysis}

Considering a rigid RN BH background, the equation of motion for a charged scalar perturbation $\delta\Phi$ can be derived by linearizing Eq.\ \eqref{eq:scalar_eq},
\begin{equation}
\left(\mathcal{D}^{\mu}\mathcal{D}_{\mu}-\mu_{\textrm{eff}}^{2}\right)\delta\Phi=0 \,,\label{eq:ScalarPert_Eq}
\end{equation}
where the effective mass is given by
\begin{equation}
\mu_{\textrm{eff}}^{2}=\alpha_{0}F^{2}/2=-\alpha_{0}Q^{2}/R^{4} \,,
\end{equation}
in terms of the RN BH charge $Q$ and the areal radius $R$. For a positive coupling constant, $\alpha_{0}>0$, the negative value of $\mu_{\textrm{eff}}^{2}$ indicates a potential tachyonic instability and the onset of spontaneous scalarization. To verify this rigorously, however, one needs to construct solutions satisfying physical boundary conditions, as we discuss next.

Focusing on a spherically symmetric perturbation, we introduce the variable $U(t,R)$ via $\delta\Phi\equiv U/R$. Eq.\ \eqref{eq:ScalarPert_Eq} then becomes
\begin{equation}
\left[\frac{\partial^{2}}{\partial t^{2}}-\frac{\partial^{2}}{\partial x^{2}}-2iqA_{t}\left(R\right)\frac{\partial}{\partial t}-q^{2}A_{t}^{2}\left(R\right)+\frac{N\left(R\right)}{R^{2}}\left(1-N\left(R\right)-\frac{Q^{2}}{R^{2}}-\alpha_{0}\frac{Q^{2}}{R^{2}}\right)\right]U\left(t,R\right)=0 \,,\label{eq:U(t,R)_Eq}
\end{equation}
where $x$ is the tortoise coordinate defined by $dx/dR=1/N\left(R\right)$, while $N\left(R\right)$ is given by
\begin{equation}
N\left(R\right)=1-\frac{2M}{R}+\frac{Q^{2}}{R^{2}} \,,
\end{equation}
where $M$ is the RN BH mass and $A_{t}\left(R\right)=-Q/R$ is the electromagnetic potential. Recall that $x=-\infty$ corresponds to the BH horizon (at $R=R_{h}$ in the areal coordinate) and $x=+\infty$ represents spatial infinity. Equivalently, in the frequency domain,
\begin{equation}
\left[\frac{\partial^{2}}{\partial x^{2}}+\left(\omega+qA_{t}\left(R\right)\right)^{2}-\frac{N\left(R\right)}{R^{2}}\left(1-N\left(R\right)-\frac{Q^{2}}{R^{2}}-\alpha_{0}\frac{Q^{2}}{R^{2}}\right)\right]U\left(R\right)=0 \,,\label{eq:U(R)_Eq}
\end{equation}
where $U(R)$ is the Fourier transform, implicitly a function of $\omega$.

A physical class of solutions is obtained by imposing the ingoing wave boundary condition at the BH horizon and the outgoing wave boundary condition at spatial infinity,
\begin{align}
U\left(R\right) & \sim e^{-ik_{h}x} & \text{as}\qquad & x\rightarrow-\infty \,,\nonumber \\
U\left(R\right) & \sim e^{+i\omega x} & \text{as}\qquad & x\rightarrow+\infty \,,\label{eq:BD_QNM}
\end{align}
where $k_{h}\equiv \omega+qA_{t}\left(R_{h}\right)$. It is noteworthy that the ingoing waves have a group velocity $v_{g}=-d\omega/dk_{h}=-1$, indicating that these waves propagate towards the horizon even though their phase velocity is positive for $k_{h}<0$. As is well known, the boundary conditions \eqref{eq:BD_QNM} select a discrete set of quasi-normal modes $\omega=\omega_{R}+i\omega_{I}$, where a positive $\omega_{I}$ indicates an unstable mode. To explore this possibility, we will numerically solve Eq. \eqref{eq:U(t,R)_Eq} for the temporal evolution of an initial Gaussian wavepacket and use the Prony method to extract the quasi-normal mode frequencies. We refer the reader to \cite{Guo:2023ivz} for details on the numerical scheme.

A different class of solutions arise by considering the scattering of an incident wave from spatial infinity. The boundary conditions in this setup read
\begin{align}
U\left(R\right) & \sim\mathcal{T}e^{-ik_{h}x} & \text{as}\qquad & x\rightarrow-\infty \,,\nonumber \\
U\left(R\right) & \sim\mathcal{I}e^{-i\omega x}+\mathcal{R}e^{+i\omega x} & \text{as}\qquad & x\rightarrow+\infty \,,\label{eq:BD_Sup}
\end{align}
where $\mathcal{I}e^{-i\omega x}$ represents the incident wave, $\mathcal{T}e^{-ik_{h}x}$ is the transmitted wave that falls into the BH horizon, and $\mathcal{R}e^{+i\omega x}$ represents the reflected wave that propagates back towards spatial infinity. It has been shown that the scattering coefficients, $\mathcal{I}$, $\mathcal{T}$ and $\mathcal{R}$, obey the relation \cite{Hod:2012wmy,Zhu:2014sya,Huang:2015cha}
\begin{equation}
\left|\mathcal{R}\right|^{2}-\left|\mathcal{I}\right|^{2}=-\frac{k_{h}}{\omega}\left|\mathcal{T}\right|^{2} \,.
\end{equation}
It follows that charged scalar waves with $0<\omega<-qA_{t}\left(R_{h}\right)$ or $-qA_{t}\left(R_{h}\right)<\omega<0$ undergo superradiant amplification, as defined by the condition $\left|\mathcal{R}\right|>\left|\mathcal{I}\right|$, resulting in the extraction of energy from the BH and the magnification of the reflected scalar wave \cite{Sanchis-Gual:2015lje,Sanchis-Gual:2016tcm,Guo:2023ivz}.

%%%%%%%%%%%%%%%%%%%%%%%%%%%%%%%%%%%
%%%%%%%%%%%%%%%%%%%%%%%%%%%%%%%%%%%

\section{Numerical Results}
\label{sec:Stable-Evolution-in}

In this section, we examine the fully nonlinear evolution of a scalar wavepacket around a RN BH, focusing first on the model without nonlinear coupling ($\alpha_{0}=0$). This scenario leads to the phenomenon of nonlinear superradiance for the charged scalar field. When a nonlinear coupling between the scalar and electromagnetic fields is introduced ($\alpha_{0}=10$ in our calculations), tachyonic instabilities trigger the formation of a scalar condensate from a small initial perturbation. Ultimately, this instability drives the system either towards a hairy BH in the case of a neutral scalar field ($q=0$) or a hairless RN BH in the charged case ($q\neq0$) through the superradiance effect.

To facilitate later analysis, we decompose the stress-energy tensor into scalar and electromagnetic components as $T_{\mu\nu}=T_{\mu\nu}^{\rm scalar}+T_{\mu\nu}^{\rm EM}$, where 
\begin{align}
T_{\mu\nu}^{\rm scalar} & =\frac{1}{8\pi}\left[\left(\mathcal{D}_{\mu}\Phi\right)^{*}\mathcal{D}_{\nu}\Phi+\left(\mathcal{D}_{\nu}\Phi\right)^{*}\mathcal{D}_{\mu}\Phi-g_{\mu\nu}\left(\mathcal{D}^{\rho}\Phi\right)^{*}\mathcal{D}_{\rho}\Phi\right] \,,\nonumber \\
T_{\mu\nu}^{\rm EM} & =\frac{1}{8\pi}\left[e^{\alpha|\Phi|^2}\left(2F_{\mu\rho}F_{\nu}^{\;\rho}-\frac{1}{2}g_{\mu\nu}F^{2}\right)\right] \,.\label{eq:T_scalar_EM}
\end{align}
This decomposition allows separation of the matter energy ($E_{\rm matter}$) and fluxes ($F_{\rm matter}^{H}$ and $F_{\rm matter}^{O}$) into scalar and electromagnetic sectors:
\begin{equation}
E_{\rm matter}=E_{\rm scalar}+E_{\rm EM}\,,\quad F_{\rm matter}^{H}=F_{\rm scalar}^{H}+F_{\rm EM}^{H}\,,\quad F_{\rm matter}^{O}=F_{\rm scalar}^{O}+F_{\rm EM}^{O}\,.
\end{equation}

It should be emphasized that, in spherically symmetric spacetimes, the electromagnetic flux vanishes due to the absence of a magnetic field, i.e.\ $F_{\rm EM}^{H}=F_{\rm EM}^{O}=0$. As a consequence, there is no spontaneous electromagnetic radiation in our setup. However, a dynamical horizon can induce an electromagnetic-like flux through the term $F_{\rm horizon}^{H}$ (see Eqs.\ \eqref{eq:horizon_flux} and \eqref{eq:T_scalar_EM}). For convenience, we denote all fluxes collectively as $F\equiv F_{\rm horizon}^{H}+F_{\rm scalar}^{H}+F_{\rm scalar}^{O}$, and the continuity equation $\left(\ref{eq:energy_contEq}\right)$ becomes 
\begin{equation}
\dot{E}_{\rm matter}=F+S \,,\label{eq:cont_Eq2}
\end{equation}
where $\dot{E}_{\rm matter}\equiv dE_{\rm matter}/dt$. The continuity equation $\left(\ref{eq:cont_Eq2}\right)$ can then be used to validate the numerical results presented below. 

In the following, we focus on an initial RN BH with charge $Q_{0}=0.9$ and mass $M_{0}=1$, which is described by 
\begin{equation}
\gamma_{ij}=W^{2}\textrm{diag}\left(1,r^{2},r^{2}\sin^{2}\theta\right)\,,\;\alpha=W^{-1}\,,\;\beta^{i}=0\,,\;E^{r}=\frac{Q_{0}}{r^{2}W^{3}}\,,\;\mathcal{A}_{\phi}=\mathcal{A}_{i}=0\,,\label{eq:wormhole_data}
\end{equation}
where $W\equiv \left(1+M_{0}/2r\right)^{2}-Q_{0}^{2}/\left(4r^{2}\right)$. A scalar wavepacket with Gaussian profile, 
\begin{equation}
\delta\Phi=pe^{-\frac{\left(r-r_{0}\right)^{2}}{2\sigma^{2}}} \,,\label{eq:Gaussian_wavepacket}
\end{equation}
is placed around the RN BH, where $r_{0}$ labels the location of the wavepacket, while $p$ and $\sigma$ characterize its amplitude and width. An observer, corresponding to the spatial boundary of our simulations, is defined in the far field at $r=R_{obs}=250M_{0}$. The evolution equation \eqref{eq:matter_eqs} for the matter fields is integrated within the CCZ3 formulation \cite{Garcia-Saenz:2025dsr} for the metric field evolution on the \textit{BlackHoles@Home} platform \cite{BHathome}. Hereafter, all numerical results are expressed in units of the initial BH mass (i.e.\ $M_{0}=1$).

%%%%%%%%%%%%%%%%%%%%%%%%%%%%%

\subsection{Nonlinear superradiance}

\begin{figure}[t]
\begin{centering}
\includegraphics[scale=0.8]{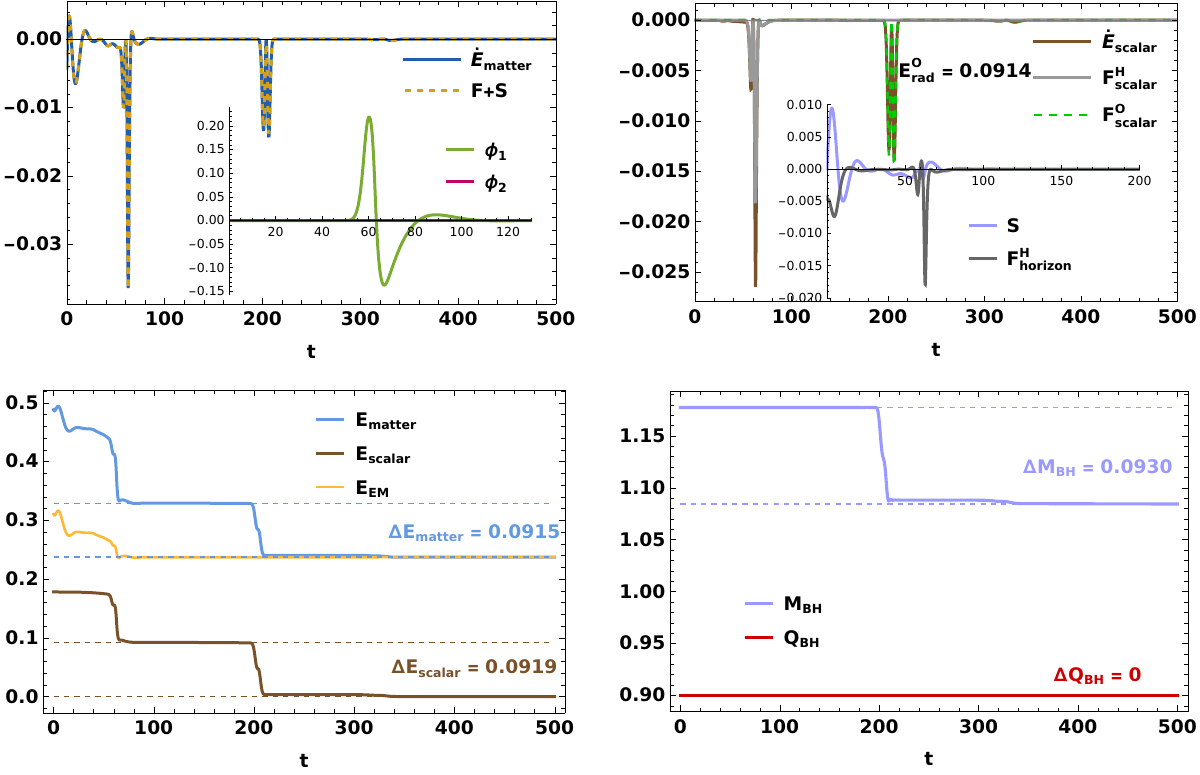}
\par\end{centering}
\caption{Numerical results for the evolution of a scalar wavepacket in a charged BH spacetime, in the context of the minimal EMS system ($\alpha_{0}=0$) with a neutral scalar field ($q=0$). {\it Top-left panel:} contributions to the energy continuity equation (cf.\ \eqref{eq:cont_Eq2}) and scalar field components. {\it Top-right:} scalar fluxes and energy rate, horizon flux and source term. {\it Bottom-left:} Total energies of the matter fields. {\it Bottom-right:} Total mass and charge of the BH spacetime. The wavepacket, initially at $r=50$, radiates approximately half of its energy into the BH at $t\approx 50$, while the other half propagates outward and is detected by an observer at $t\approx200$. A weak reflection, resulting from scattering between the incident waves and the curvature, is detected by the observer at $t\approx300$. The total scalar radiation measured is $E_{\rm rad} = 0.0914$, with no charge transport.}
\label{EM_nonPert}
\end{figure}
\begin{figure}[t]
\begin{centering}
\includegraphics[scale=0.8]{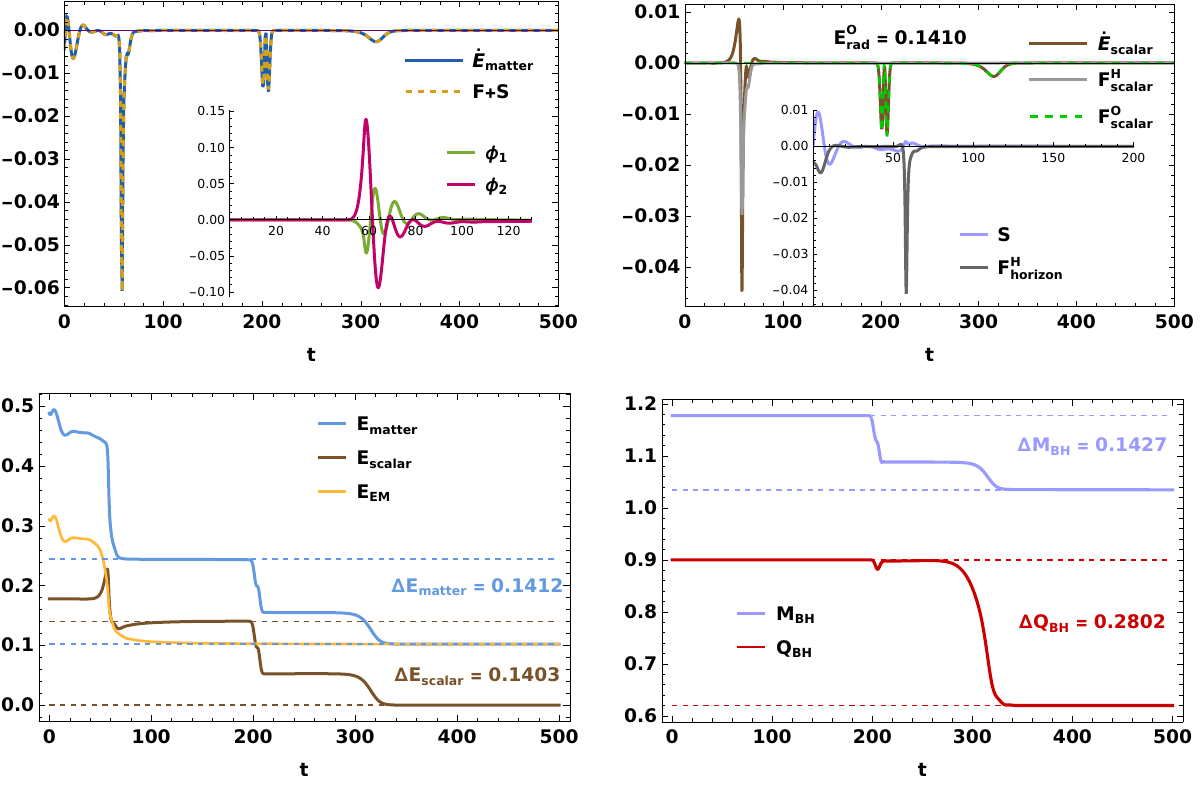}
\par\end{centering}
\caption{Numerical results for the evolution of a scalar wavepacket in a charged BH spacetime, in the context of the minimal EMS system ($\alpha_{0}=0$) with a charged scalar field ($q=0.4$). The panels match those of Fig.\ \ref{EM_nonPert}. High-frequency components of the wavepacket are absorbed by the BH, while low-frequency waves undergo superradiant amplification, reflected to infinity and detected at $t\approx300$. The results confirm that superradiance takes place through the energy transfer from the electromagnetic field to the scalar field, and demonstrate the process of extraction of BH charge and energy by the scalar field, carried
to spatial infinity.}
\label{EM_Superradiance}
\end{figure}

We first study the standard setup for the superradiant amplification of a scalar wave by a RN BH, where the scalar field is only minimally coupled with both the gravitational and the electromagnetic fields within the EMS model ($\alpha_{0}=0$). We refer to this effect as ``nonlinear superradiance'', to emphasize that our treatment is fully nonlinear, and also to distinguish it from the scalarization model (i.e.\ with nonminimal electromagnetic coupling) that we investigate in the next subsection.

The scalar wavepacket is described by Eq.\ \eqref{eq:Gaussian_wavepacket} with the choice of parameters $p=0.02$ and $\sigma=5$, and is initially placed at $r_{0}=50$. In this setup, the system has the total energy $M=1.1789$, consisting of the background BH energy $M_{0}=1$ and the wavepacket energy $M_{\rm scalar}=0.1789$. Details on solving for the initial BH data can be found in Appendix \ref{sec:appdA}.

The initial BH, described by the wormhole-like data \eqref{eq:wormhole_data} or \eqref{eq:ansatz}, begins in a non-equilibrium state and evolves into a trumpet slice \cite{Campanelli:2005dd,Baker:2005vv,Brown:2009ki}. From our simulations, we extract several physical quantities as functions of time, including the energies and fluxes for the different matter sources; the scalar field components $\phi_1$ and $\phi_2$, defined via $\Phi=\phi_1+i\phi_2$; and total energy and charge of the system as given by the quantities $M_{\rm BH}$ and $Q_{\rm BH}$ discussed in the previous section.

Fig.\ \ref{EM_nonPert} presents the results in the case of a neutral scalar field ($q=0$). The non-equilibrium nature of the initial configuration is made clear by the time evolution, notably by the initial negative value of $F_{\rm horizon}^{H}$ (upper-right panel), indicating horizon expansion. The upper-left panel illustrates the perfect agreement (within numerical precision) of our simulation with the continuity equation for energy \eqref{eq:cont_Eq2}; the inset displays the scalar field components, and we observe that the imaginary direction $\phi_2$ is not excited since we choose its initial condition to be zero and we observe that $\phi_1$ and $\phi_2$ are decoupled when $q=0$, cf.\ \eqref{eq:scalar_eq}. The bottom-left panel highlights in particular the evolution of the scalar field's energy (brown curve): one observes that the field radiates approximately half of its energy into the BH horizon $t\approx r_{0}=50$, i.e.\ the light-crossing time between the horizon and the initial wavepacket, while the other half is emitted to spatial infinity, where it is detected by the observer at $t\approx R_{obs}-r_{0}=200$. This ``equipartition'' is expected from the fact that the wavepacket is chosen with zero initial momentum in our simulations. Finally, a small fraction of the scalar radiation interacts with the BH's potential, resulting in a weak reflected scalar wave which is detected by the observer at $t\approx r_{0}+R_{obs}=300$.

The processes of radiation into the horizon and towards the spatial boundary are confirmed by a calculation of the flux components $F^H_{\rm scalar}$ and $F^O_{\rm scalar}$, see in particular the gray and green curves in the upper-right panel of Fig.\ \ref{EM_nonPert}. The total scalar radiation measured by the observer is $E_{\rm rad}^{O}=\int_{0}^{500}\left(-F_{\rm scalar}^{O}\right)dt=0.0914$, which includes the initial radiation from the wavepacket, $\int_{0}^{250}\left(-F_{\rm scalar}^{O}\right)dt=0.0882\approx M_{\rm scalar}/2$, and the radiation from the reflected wave $\int_{250}^{500}\left(-F_{\rm scalar}^{O}\right)dt=0.0032$, where $t=250$ is here chosen as a suitable middle point between the times that characterize the two effects. Finally, the bottom-right panel displays the time evolution of the BH mass and charge as determined by the fields at the location of the observer, i.e.\ the expressions in Eqs.\ \eqref{eq:BHmass} and \eqref{eq:BHcharge}. The numerical results confirm that there is no charge transport in the absence of electric charge in the scalar field. 

%----------------------

In Fig.\ \ref{EM_Superradiance} we present the numerical results for the same setup as above, except that the scalar field is now charged, with $q=0.4$. According to the discussion of Sec.\ \ref{subsec:Linear-analysis}, we expect qualitatively new features due to the fact that scalar waves with frequencies in the range $0<\omega<-qA_{t}\left(R_{h}\right)$ can undergo superradiant amplification. 

The upper-left panel qualitatively illustrates the phenomenon, showing that high-frequency components of the wavepacket penetrate the curvature potential barrier and are absorbed by the BH (see the inset), while low-frequency scalar waves are reflected towards spatial infinity, forming a significant signal detected by the observer at the light-crossing time $t\approx r_{0}+R_{obs}=300$. The upper-right panel evinces the presence of a positive scalar current $\dot{E}_{\rm scalar}$ when the wavepacket starts scattering off the curvature at $t\approx r_{0}=50$, indicating the superradiance effect. As a result of this process, the total scalar radiation detected by the observer is $E_{\rm rad}^{O}=\int_{0}^{500}\left(-F_{\rm scalar}^{O}\right)dt=0.1410$, which includes the initial radiation, $\int_{0}^{250}\left(-F_{\rm scalar}^{O}\right)dt=0.0882\approx M_{\rm scalar}/2$, and the amplified reflected radiation $\int_{250}^{500}\left(-F_{scalar}^{O}\right)dt=0.0528$. 

Interestingly, the bottom-left panel illustrates the mechanism behind superradiance: the energy gain of the scalar waves originates from the transfer of energy from the electromagnetic field to the scalar field, even as the total matter energy decreases. The bottom-right panel further shows that, in the presence of charge in the scalar field, the nonlinear superradiance of scalar waves leads to the extraction of both BH energy and charge, which are then carried to spatial infinity.

%%%%%%%%%%%%%%%%%%%%%%%%%%%%

\subsection{Tachyonic superradiance}

\begin{figure}[t]
\begin{centering}
\includegraphics[scale=0.8]{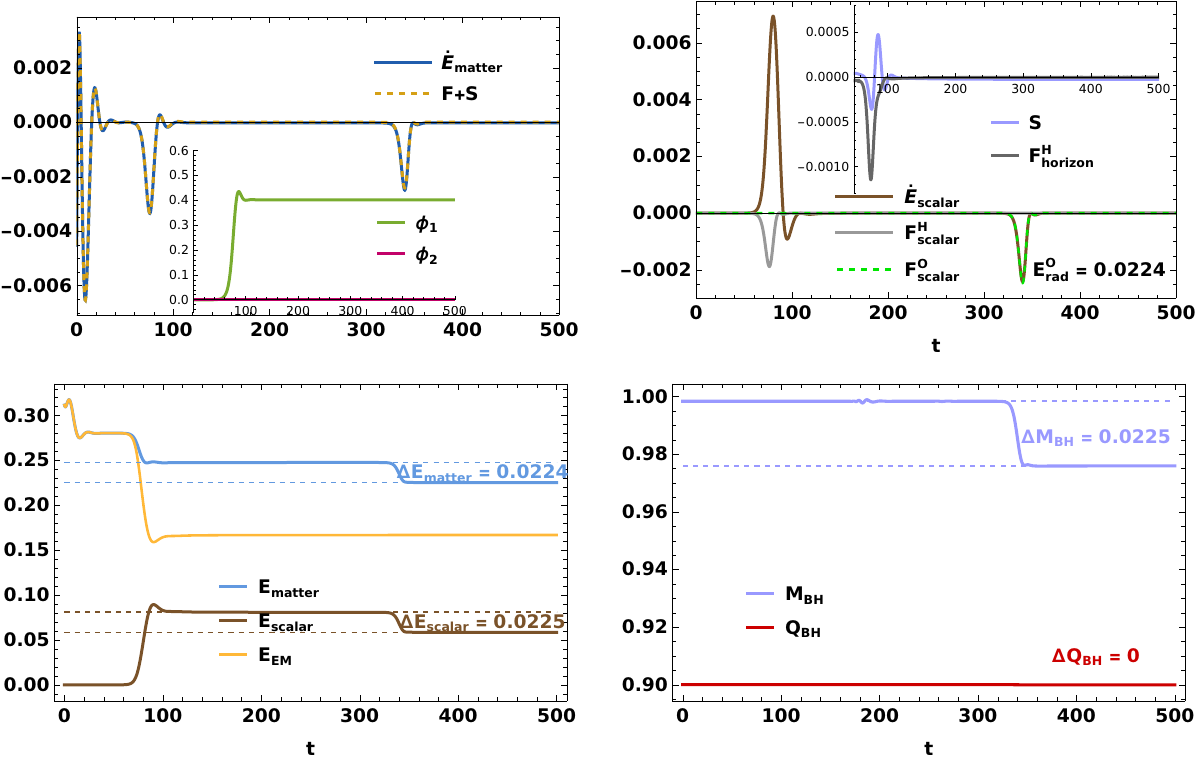}
\par\end{centering}
\caption{Numerical results for the evolution of a scalar wavepacket in a charged BH spacetime, in the context of the nonminimal EMS system ($\alpha_{0}=10$) with a neutral scalar field ($q=0$). The panels match those of Fig.\ \ref{EM_nonPert}. The initial scalar perturbation triggers tachyonic instabilities, leading to the formation of a hairy BH via spontaneous scalarization (upper-left panel). Energy is transferred from the electromagnetic field to the scalar field during scalar condensation (bottom-left panel), while part of the matter energy is absorbed by the BH as shown by the horizon expansion (upper-right panel). The scalar cloud relaxes, radiating waves to spatial infinity and returning a small fraction of energy to the electromagnetic field (lower-left panel). The final state is a lower-energy equilibrium hairy BH, with no charge transport observed (bottom-right panel).}
\label{EMS_scalarization}
\end{figure}
\begin{figure}[t]
\begin{centering}
\includegraphics[scale=0.8]{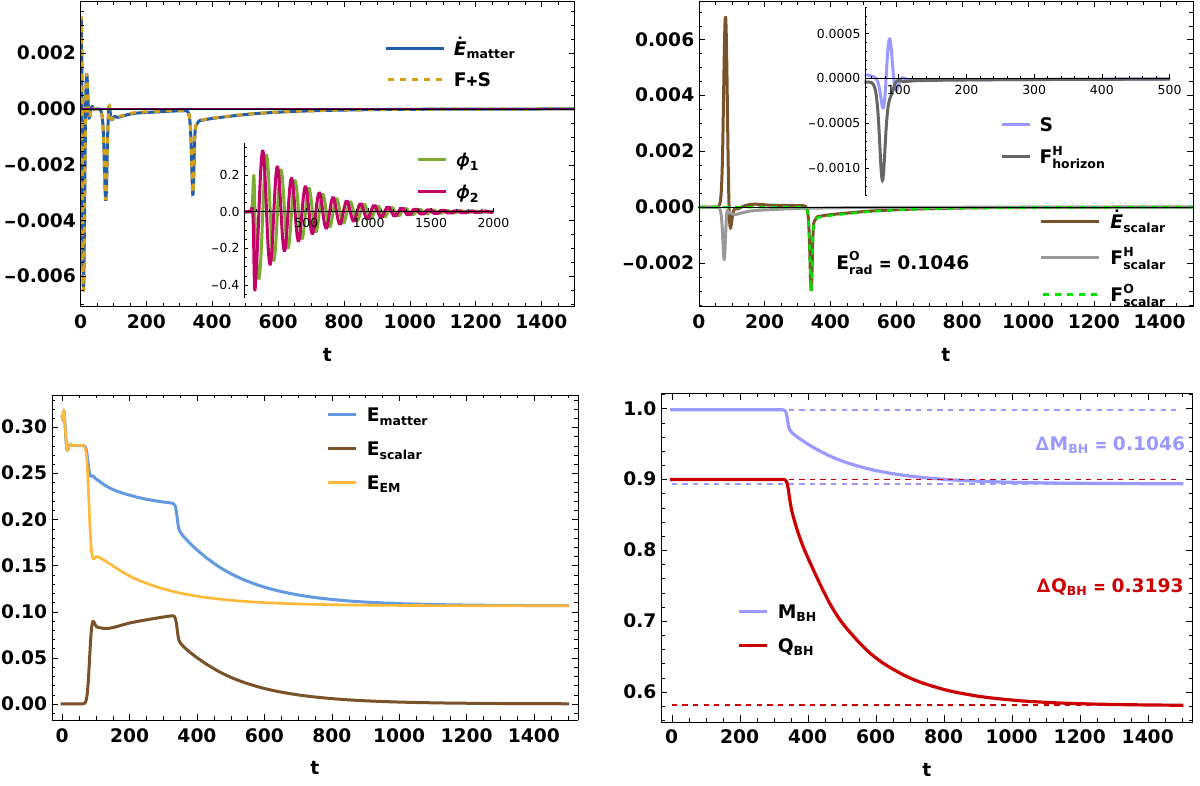}
\par\end{centering}
\caption{Numerical results for the evolution of a scalar wavepacket in a charged BH spacetime, in the context of the nonminimal EMS system ($\alpha_{0}=10$) with a charged scalar field ($q=0.2$). The panels match those of Fig.\ \ref{EM_nonPert}. The simulation demonstrates the competing effects of scalar condensation and superradiance. A positive scalar current $\dot{E}_{\rm scalar}$ from superradiance is measured after cloud formation (around $100<t<300$), despite the loss of scalar energy via BH accretion (negative $F_{\rm scalar}^{H}$). The effect is also illustrated by the energy extraction from the electromagnetic field, increasing the scalar energy while decreasing the total matter energy, along with the extraction of BH energy and charge by scalar radiation into spatial infinity.}
\label{EMS_q005}
\end{figure}
\begin{figure}[t]
\begin{centering}
\includegraphics[scale=0.8]{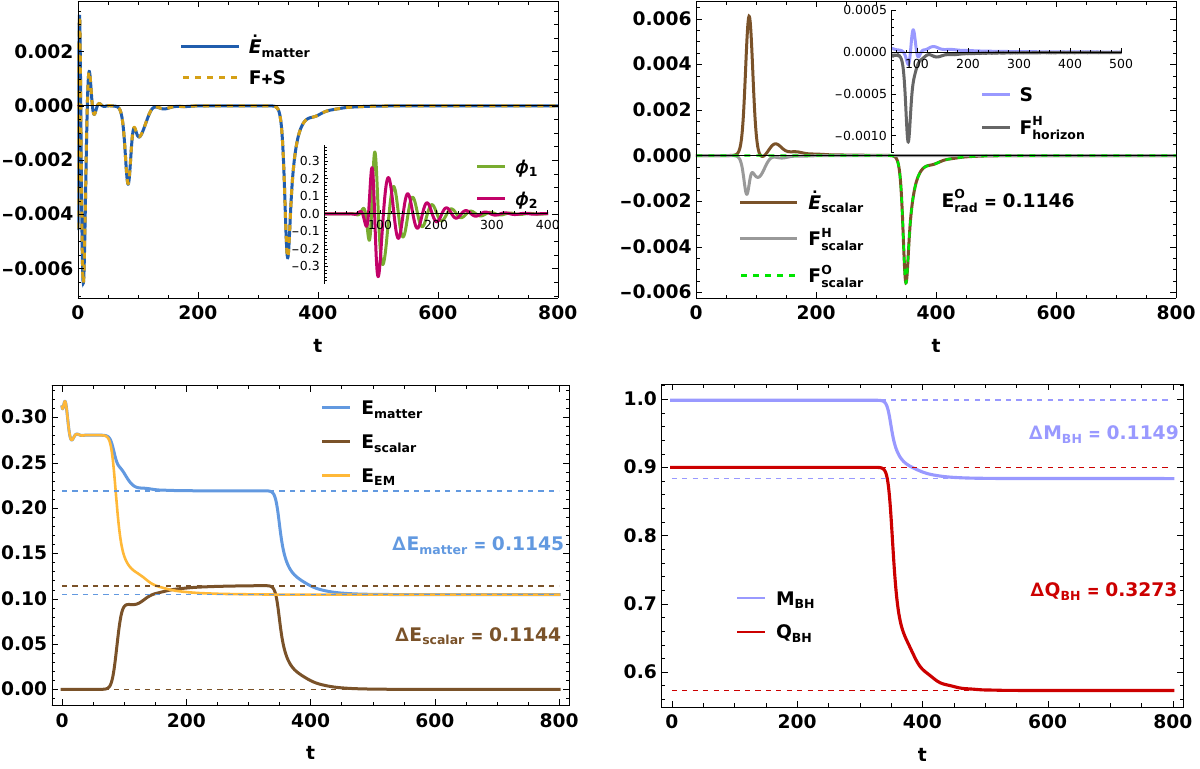}
\par\end{centering}
\caption{Numerical results for the evolution of a scalar wavepacket in a charged BH spacetime, in the context of the nonminimal EMS system ($\alpha_{0}=10$) with a charged scalar field ($q=0.6$). The panels match those of Fig.\ \ref{EM_nonPert}. Larger values of the electric charge $q$ more clearly illustrate the tachyonic superradiance effect. The scalar condensate rapidly converts into radiation, extracting BH energy and charge toward spatial infinity. The final state is again a hairless RN BH free of tachyonic instabilities.}
\label{EMS_q015}
\end{figure}

Next, we turn our attention to the nonlinear evolution of a scalar perturbation in a charged BH spacetime described by the EMS model, with the nonminimal coupling constant set to $\alpha_{0}=10$. The initial scalar wavepacket is described by \eqref{eq:Gaussian_wavepacket} with parameters $p=10^{-6}$, $\sigma=5$ and $r_{0}=5$. Owing to the negligible amplitude of the perturbation, the evolution of the system maintains the total energy and charge within numerical precision, i.e.\ $M\approx M_{0}=1$, and $Q\approx Q_{0}=0.9$. As in the previous subsection, we study both the uncharged and charged scalar field cases, as a way to better understand the effects of electric charge in the evolution of matter fields in this theory. We expect an interesting interplay of both tachyonic and superradiant effects in the general case, which we refer to as ``tachyonic superradiance''.

The existence of a tachyonic instability may be verified by solving Eq.\ \eqref{eq:U(R)_Eq} with the boundary condition \eqref{eq:BD_QNM} for the dominant unstable mode and its frequency $\omega_0$ (where by `dominant' we mean the mode with greatest imaginary part\footnote{Actually, for the parameter settings we have studied, only a single unstable mode exists in each case. Higher (less dominant) unstable overtones will be present if the depth of the effective potential in Eq.\ \eqref{eq:U(R)_Eq} is larger, in particular if $Q$ or $\alpha_0$ are larger.}). We display the results in Table \ref{table} for three different values of the scalar field charge $q$. Notice that unstable modes are not characterized by a purely imaginary frequency in the charged case, owing to the complex nature of the quasi-normal mode equation.

Fig.\ \ref{EMS_scalarization} presents the results of our numerical simulation in the model with uncharged scalar field ($q=0$), i.e.\ the setup already considered previously in \cite{Garcia-Saenz:2024beb}. The figure demonstrates the dynamics of spontaneous scalarization initiated by a neutral scalar field, revealing the formation of a hairy BH induced by the tachyonic growth of the field; see in particular the inset in the upper-left panel. During this process, the development of a massive scalar condensate results in the transfer of energy from the electromagnetic field to the scalar field (lower-left panel). Simultaneously, a portion of the matter energy is absorbed by the BH due to the expanding horizon, as evidenced by the negative horizon flux (upper-right panel). Eventually, the scalar condensation undergoes relaxation, returning a small fraction of energy to the electromagnetic field (lower-left panel) and radiating scalar waves ($E_{\rm rad}^{O}=0.0224$) towards spatial infinity (upper-right panel). The final state is a lower-energy equilibrium hairy BH. The lower-right panel confirms that no BH charge transport occurs during spontaneous scalarization by a neutral scalar field. 

In Table \ref{table}, the charge-to-mass ratio of the final hairy BH (bottom two rows) is obtained through two approaches: (1) by calculating the energy and charge transport of the BH from the simulations, as shown e.g.\ in Fig.\ \ref{EMS_scalarization}, and (2) by solving for the hairy BH solution that matches the scalar field value at the horizon (see Appendix \ref{sec:appB}). Both methods yield consistent results for the final $Q/M$ of the spontaneous scalarization process, incidentally providing a nontrivial check of our numerics.

%------------

We consider at last the general setup with charged scalar field and nonminimal electromagnetic coupling in the EMS model. We study first the case of small electric charge, setting $q=0.2$. As already mentioned, the quasi-normal mode spectrum still features an unstable mode (in this case $\omega_{0}=0.0969+0.1811i$), inducing the formation of a scalar cloud, similarly to the scalarization process observed in Fig.\ \ref{EMS_scalarization}. However, the field also experiences now the superradiant effect, and the expectation from linear analysis is that an increased amount of energy will be emitted towards spatial infinity, potentially causing the depletion of the scalar condensate.

Our numerical simulation, shown in Fig.\ \ref{EMS_q005}, confirms that superradiance results in the full depletion of the scalar cloud formed as a consequence of the tachyonic destabilization. The top-right panel illustrates both effects, as well as their qualitative differences, as seen in particular in the curve of $\dot{E}_{\rm scalar}$ (brown): the large peak characteristic of spontaneous scalarization (cf.\ Fig.\ \ref{EMS_scalarization}) is followed by a period of net growth of the total energy of the scalar, despite a negative flux $F_{\rm scalar}^{H}$ that signifies the absorption of scalar radiation by the BH during this process. The two phases are also clear in the bottom-left panel in the figure, which further highlights the mechanism of superradiance, namely the amplification of scalar waves via the extraction of BH energy through the electromagnetic sector. Finally, and more importantly, the panel makes evident that superradiant scattering results in the complete depletion of the scalar cloud, as is also legible in the upper-left panel (inset) showing the damping of the scalar field, and in the upper-right panel showing the persistence of scalar radiation detected by the observer in the far-field region.

It is known that tachyonic modes exist in the EMS model only if the charge-to-mass ratio ($Q/M$) of the BH exceeds a certain critical value, for a given coupling $\alpha_0$ \cite{Herdeiro:2018wub}. We refer to this value as the bifurcation point.\footnote{The bifurcation point is numerically determined by solving Eq.\ \eqref{eq:U(R)_Eq} for the quasi-normal mode frequency with vanishing imaginary part, which marks the critical value of $Q/M$.} We find that the bifurcation point is also a function of the charge $q$ of the scalar field, and appears to increase with $q$, at least in the small-$q$ range that we have explored; cf.\ the second row of Table \ref{table}. We find it intriguing that superradiance tends to alleviate the tachyonic effect, as a larger charge $q$ appears to slightly decrease the imaginary part of the initial unstable mode or increase the bifurcation point $Q/M$. Most notably, Table \ref{table} reveals that the final state of evolution possesses a slightly lower $Q/M$ compared to the bifurcation point, ensuring that the final state remains a hairless RN BH free from tachyonic instabilities.

%-------------

The results of the simulation for a larger value of the charge, $q=0.6$, are shown in Fig.\ \ref{EMS_q015}. The numerical results show a similar evolution to that discussed in Fig.\ \ref{EMS_q005}, but the increased scalar field charge is seen to accelerate the superradiance process driven by the scalar condensate. As shown in the bottom-left panel, the cloud fully converts into scalar radiation via superradiance, reaching a maximum energy plateau before scalar waves are detected by the distant observer. Eventually, this significant scalar radiation extracts energy and charge from the BH toward spatial infinity, leaving behind a hairless RN BH without further instabilities.

\begin{table}
\begin{centering}
\includegraphics[scale=0.25]{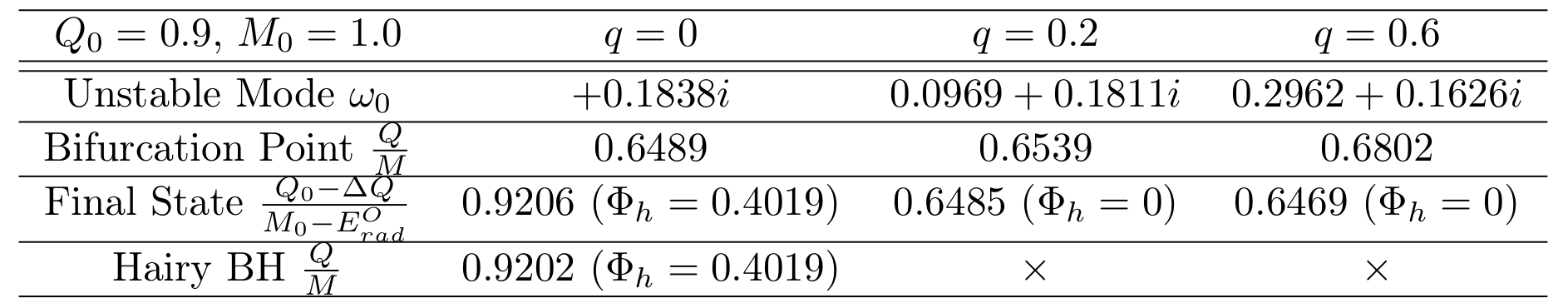}
\par\end{centering}
\caption{Numerical results for the evolution of a scalar perturbation in the EMS system ($\alpha_{0}=10$) around an initial RN BH ($Q_{0}=0.9$, $M_{0}=1$). The unstable modes with frequency $\omega_{0}$ characterize the initial tachyonic instability. The bifurcation point $Q/M$ represents the critical ratio above which tachyonic instabilities emerge. The final state corresponds to the endpoint of the nonlinear evolution driven by these instabilities. The hairy BH solution is obtained numerically by solving the EMS equations with a static ansatz, for a given value $\Phi_h$ of the scalar field at the horizon (see Appendix \ref{sec:appB}). One observes the good agreement of the values of $Q/M$ extracted from the dynamical simulation and from the analysis of the static system, validating the accuracy of the numerical calculations.}
\label{table}
\end{table}

\begin{figure}[t]
\begin{centering}
\includegraphics[scale=0.8]{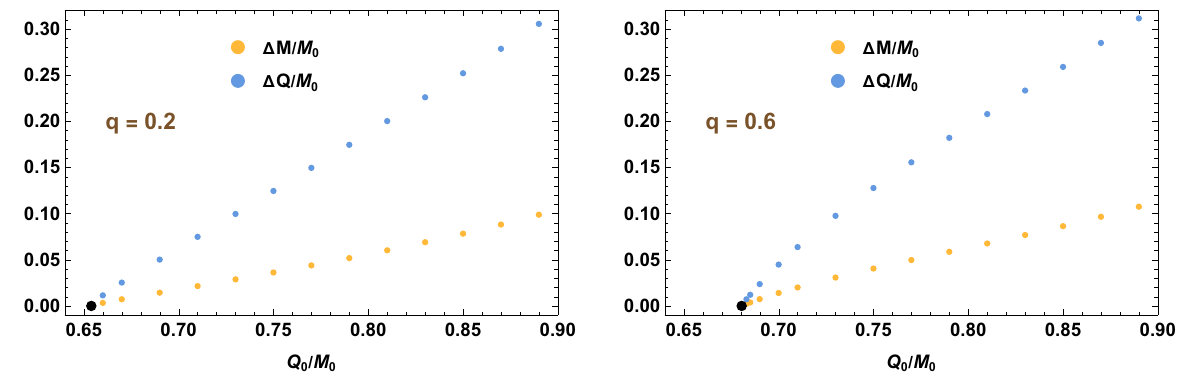}
\par\end{centering}
\caption{Numerical results for the energy and charge loss fractions in the EMS system ($\alpha_{0}=10$), plotted as functions of the charge-to-mass ratio $Q_0/M_0$ of the initial RN BH, using $q=0.2$ (left panel) and $q=0.6$ (right). The energy loss $\Delta M$ is computed from the scalar radiation $E_{\rm rad}$. The black circles indicate the bifurcation points quoted in Table \ref{table}, i.e.\ the critical points for the onset of tachyonic quasi-normal modes in the linear approximation.}
\label{fig:bifurcation}
\end{figure}

Throughout this section we have considered the fixed value $Q_0/M_0=0.9$ for the charge-to-mass ratio of the initial RN BH. It is also interesting to study how changing this ratio affects observables. In Fig.\ \ref{fig:bifurcation} we display the dependence of the fractional losses of energy and charge as functions of $Q_0/M_0$. We observe a monotonic dependence of the loss fractions as the charge-to-mass ratio increases, consistent with the expectation that a higher destabilization rate should lead to greater amounts of scalar radiation. Our results also show the convergence of the curves towards the bifurcation point calculated in linear theory, providing a non-trivial consistency check of our numerical calculations.

%%%%%%%%%%%%%%%%%%%%%%%%%%%%
%%%%%%%%%%%%%%%%%%%%%%%%%%%%

\section{Conclusions}
\label{Sec:Conc}

Our aim in this paper was to provide, through fully nonlinear numerical simulations, a comprehensive picture of BH dynamics in the presence of tachyonic instabilities and superradiant amplification of scalar fields. The two effects collaborate to create a macroscopic scalar condensate out of an initial small perturbation, yet the two processes remain distinguishable in the time evolution, as evidenced by our detailed results on the time dependence of the different energy fluxes that characterize the system. One of our main conclusions is that superradiance results in the complete depletion of the scalar cloud, with a significant fraction of energy and charge being lost into radiation toward spatial infinity. Our energetics analysis has also revealed that the superradiant amplification of charged scalar waves results, as expected, from the extraction of electromagnetic field energy from the BH.

We take our results as a strong indication that stable, static and spherically symmetric BH solutions endowed with scalar hair do not exist in the EMS system in the presence of electric charge. While our numerical simulations clearly suggest this, we also have additional evidence from the study of the EMS equations with a static ansatz, cf.\ Appendix \ref{sec:appB}. Upon performing an extensive scan of the parameter space, we have been unable to find scalarized solutions with $q\neq 0$. On the other hand, to our knowledge, the absence of hairy BH solutions in the charged EMS model does not follow from any existing no-hair theorem (see \cite{Herdeiro:2015waa} for a review), or at least not as an immediate corollary thereof. In particular, the pioneering work \cite{Mayo:1996mv} does not encompass nonminimal couplings between scalar and electromagnetic fields; see also \cite{Hong:2020miv,Herdeiro:2020xmb,Hartnoll:2020fhc,Cai:2020wrp,An:2023bpb,Herdeiro:2024yqa} for related results. It is natural to expect, however, that generalizations of such theorems should exist that accommodate nontrivial couplings between scalar fields and gauge fields, which would allow one to analytically demonstrate the absence of hairy solutions in the charged EMS model, at least under the assumption of spherical symmetry. We plan to address this question in future work.

While the end state of the evolution in the presence of the superradiant effect is the same with and without the nonminimal coupling, the details of the evolution are clearly affected by the scalarization process. Both the total amount and flux of scalar energy radiated depend on the initial tachyonic growth of the scalar field, which is therefore observable. The tachyonic superradiance effect could also serve as a mechanism to generate sizeable amounts of scalar radiation from small initial perturbations, which we have claimed to be particularly efficient as it results in the total depletion of the scalar cloud. It would be interesting to extend our study to the situation without spherical symmetry, thus opening the channels for electromagnetic and gravitational radiation. On the other hand, as we alluded to in the introduction, the physical relevance of this mechanism is questionable due to the need of significant BH charge. However, we expect that the interesting interplay between scalarization and superradiance will persist in other systems which do not rely on BH charge, such as a rotating BH in scalar-Gauss-Bonnet theory.

\begin{acknowledgments}
We are grateful to Yiqian Chen for useful discussions and valuable comments. SGS, GG and XW are supported by the NSFC (Grant Nos. 12250410250 and 12347133). SGS also acknowledges support from a Provincial Grant (Grant No.\ 2023QN10X389). PW is supported in part by the NSFC (Grant Nos. 12105191, 12275183, 12275184 and 11875196).
\end{acknowledgments}

\appendix

\section{Initial data}
\label{sec:appdA}

In this appendix, we describe the equations needed to determine the initial BH data after placing a non-perturbative scalar wavepacket around a RN BH. The constraint
equations for the metric field are given by 
\begin{eqnarray}
\mathcal{H} & = & ^{(3)}R-K_{ij}K^{ij}+K^{2}-16\pi\rho=0 \,,\nonumber \\
\mathcal{M}_{i} & = & D_{j}K_{i}^{j}-D_{i}K-8\pi S_{i}=0 \,,\label{eq:constraint_metric_eqs}
\end{eqnarray}
where $^{(3)}R$ is the 3-dimensional curvature scalar associated with the metric $\gamma_{ij}$. By projecting the Maxwell equation $\left(\ref{eq:A_eq}\right)$ onto the normal vector $n_{\mu}$, one can obtain the constraint equations for the electromagnetic field,
\begin{equation}
D_{i}E^{i}=4\pi\rho_{e} \,,\quad D_{i}B^{i}=0 \,.\label{eq:constraint_EM_eqs}
\end{equation}

To solve Eqs.\ \eqref{eq:constraint_metric_eqs} and \eqref{eq:constraint_EM_eqs} in spherically symmetric spacetimes, we consider the following ansatz:
\begin{equation}
\gamma_{ij}=W^{2}\left(r\right)\textrm{diag}\left(1,r^{2},r^{2}\sin^{2}\theta\right)\,,\;\alpha=W^{-1}\left(r\right)\,,\;\Phi=\Phi\left(r\right)\,,\;\beta^{i}=K_{ij}=\mathcal{A}_{*}=\mathcal{A}_{i}=\Pi=0\,.\label{eq:ansatz}
\end{equation}
Substituting this into the above constraints one can obtain the equation
\begin{equation}
W^{\prime\prime}\left(r\right)+\frac{2}{r}W^{\prime}\left(r\right)-\frac{1}{2W\left(r\right)}W^{\prime}\left(r\right)^{2}+\frac{1}{2}W\left(r\right)\Phi^{\prime}\left(r\right)^{2}+\frac{Q^{2}}{2r^{4}W\left(r\right)f\left(\Phi\right)}=0 \,.\label{eq:W_eq}
\end{equation}
Notice that the electric field is determined as $E^{r}=Q/\left(W\left(r\right)^{3}r^{2}f\left(\Phi\right)\right)$ in terms of the function $W$, where $Q$ is the BH charge and primes denote derivatives with respect to $r$. Additionally, the BH horizon $r_{h}$ is determined by 
\begin{equation}
W^{\prime}\left(r_{h}\right)=-\frac{W\left(r_{h}\right)}{r_{h}} \,,
\end{equation}
which provides the boundary condition, together with $W\left(r=\infty\right)=1$, for solving Eq.\ \eqref{eq:W_eq} for the function $W\left(r\right)$. 

Considering a spherically symmetric Gaussian wavepacket as the initial configuration for the scalar field, placed around a RN BH of mass $M_{0}$, charge $Q_{0}$ and horizon radius $r=r_{h}$, the metric function $W\left(r\right)$ is obtained by solving Eq.\ \eqref{eq:W_eq} with fixed values of $Q_{0}$ and $r_{h}$. The system then has the total energy (from Eq.\ \eqref{eq:dl2})
\begin{equation}
M=\lim_{r\rightarrow\infty}\left[-r^{2}W\left(r\right)^{2}W^{\prime}\left(r\right)\right] \,,
\end{equation}
which includes both the initial BH mass $M_{0}$ and the additional contribution from the scalar field.

%%%%%%%%%%%%%%%%%%%%%%%%%%%

\section{Static hairy black hole solution}
\label{sec:appB}

As is well known, and as demonstrated in this paper, the EMS model with neutral scalar field accommodates static hairy BH solutions. These are determined by numerically solving the Einstein and matter equations described in Sec.\ \ref{secII}. To this end we consider the following spherically symmetric ansatz:
\begin{align}
ds^{2} & =-N(R)e^{-2\delta(R)}dt^{2}+\frac{1}{N(R)}dR^{2}+R^{2}\left(d\theta^{2}+\sin^{2}\theta d\varphi^{2}\right) \,,\nonumber \\
A_{\mu}dx^{\mu} & =A_{t}(R)dt \,,\quad \Phi=\phi(R) \,,\label{eq:HBH_ansatz}
\end{align}
respectively for the metric, electromagnetic and scalar fields. Note that, in the neutral case, the scalar field may be assumed real without loss of generality. The above ansatz leads to the following set of ordinary differential equations:
\begin{align}
N^{\prime}(R) & =\frac{1-N(R)}{R}-\frac{Q^{2}}{R^{3}e^{\alpha_{0}\phi^{2}(R)}}-RN(R)\left[\phi^{\prime}(R)\right]^{2} \,,\nonumber \\
\left[R^{2}N(R)\phi^{\prime}(R)\right]^{\prime} & =-\frac{\alpha_{0}Q^{2}\phi(R)}{R^{2}e^{\alpha_{0}\phi^{2}(R)}}-R^{3}N(R)\left[\phi^{\prime}(R)\right]^{3} \,,\nonumber \\
\delta^{\prime}(R) & =-R\left[\phi^{\prime}(R)\right]^{2} \,,\label{eq:EOM}\\
A_{t}^{\prime}(R) & =\frac{Q}{R^{2}e^{\alpha_{0}\phi^{2}(R)}}e^{-\delta(R)} \,,\nonumber 
\end{align}
where $Q$ is the BH charge.

To solve for hairy BH states one must impose appropriate boundary conditions. At the event horizon $R_{h}$, one has
\begin{equation}
N(R_{h})=0\,,\; \delta(R_{h})=\delta_{0} \,,\; \phi(R_{h})=\phi_{0} \,,\; A_{t}(R{}_{h})=V_{0} \,,\label{eq:Rh_condition}
\end{equation}
where $V_{0}$ is the horizon electrostatic potential, and $\delta_{0}$ and $\phi_{0}$ are parameters that characterize a specific BH solution. At spatial infinity, the asymptotic behavior is
\begin{equation}
N(R)=1-\frac{2M}{R}+\frac{Q^{2}+Q_{s}^{2}}{R^{2}}+\ldots \,,\; \delta(R)=\frac{Q_{s}^{2}}{2R^{2}}+\ldots \,,\; \phi(R)=\frac{Q_{s}}{R}+\ldots \,,\; A_{t}(R)=-\frac{Q}{R}+\ldots \,, \label{eq:infinity_condition}
\end{equation}
where $M$ is the BH mass, and $Q_{s}$ is the scalar charge (not to be confused with the electric charge of the scalar field). For given values of $\alpha_{0}$ and $\phi_{0}$, one can numerically solve Eqs.\ \eqref{eq:EOM} for the BH configuration that satisfies the above boundary conditions.

\bibliographystyle{unsrturl}
\bibliography{ref}

\begin{thebibliography}{10}

\bibitem{Arvanitaki:2009fg}
Asimina Arvanitaki, Savas Dimopoulos, Sergei Dubovsky, Nemanja Kaloper, and
  John March-Russell.
\newblock {String Axiverse}.
\newblock {\em Phys. Rev. D}, 81:123530, 2010.
\newblock \href {http://arxiv.org/abs/0905.4720} {\path{arXiv:0905.4720}},
  \href {https://doi.org/10.1103/PhysRevD.81.123530}
  {\path{doi:10.1103/PhysRevD.81.123530}}.

\bibitem{Arvanitaki:2010sy}
Asimina Arvanitaki and Sergei Dubovsky.
\newblock {Exploring the String Axiverse with Precision Black Hole Physics}.
\newblock {\em Phys. Rev. D}, 83:044026, 2011.
\newblock \href {http://arxiv.org/abs/1004.3558} {\path{arXiv:1004.3558}},
  \href {https://doi.org/10.1103/PhysRevD.83.044026}
  {\path{doi:10.1103/PhysRevD.83.044026}}.

\bibitem{Brito:2014wla}
Richard Brito, Vitor Cardoso, and Paolo Pani.
\newblock {Black holes as particle detectors: evolution of superradiant
  instabilities}.
\newblock {\em Class. Quant. Grav.}, 32(13):134001, 2015.
\newblock \href {http://arxiv.org/abs/1411.0686} {\path{arXiv:1411.0686}},
  \href {https://doi.org/10.1088/0264-9381/32/13/134001}
  {\path{doi:10.1088/0264-9381/32/13/134001}}.

\bibitem{Ferreira:2017pth}
Miguel~C. Ferreira, Caio F.~B. Macedo, and Vitor Cardoso.
\newblock {Orbital fingerprints of ultralight scalar fields around black
  holes}.
\newblock {\em Phys. Rev. D}, 96(8):083017, 2017.
\newblock \href {http://arxiv.org/abs/1710.00830} {\path{arXiv:1710.00830}},
  \href {https://doi.org/10.1103/PhysRevD.96.083017}
  {\path{doi:10.1103/PhysRevD.96.083017}}.

\bibitem{Baumann:2018vus}
Daniel Baumann, Horng~Sheng Chia, and Rafael~A. Porto.
\newblock {Probing Ultralight Bosons with Binary Black Holes}.
\newblock {\em Phys. Rev. D}, 99(4):044001, 2019.
\newblock \href {http://arxiv.org/abs/1804.03208} {\path{arXiv:1804.03208}},
  \href {https://doi.org/10.1103/PhysRevD.99.044001}
  {\path{doi:10.1103/PhysRevD.99.044001}}.

\bibitem{Zhang:2019eid}
Jun Zhang and Huan Yang.
\newblock {Dynamic Signatures of Black Hole Binaries with Superradiant Clouds}.
\newblock {\em Phys. Rev. D}, 101(4):043020, 2020.
\newblock \href {http://arxiv.org/abs/1907.13582} {\path{arXiv:1907.13582}},
  \href {https://doi.org/10.1103/PhysRevD.101.043020}
  {\path{doi:10.1103/PhysRevD.101.043020}}.

\bibitem{Garcia-Saenz:2021uyv}
Sebastian Garcia-Saenz, Aaron Held, and Jun Zhang.
\newblock {Destabilization of Black Holes and Stars by Generalized Proca
  Fields}.
\newblock {\em Phys. Rev. Lett.}, 127(13):131104, 2021.
\newblock \href {http://arxiv.org/abs/2104.08049} {\path{arXiv:2104.08049}},
  \href {https://doi.org/10.1103/PhysRevLett.127.131104}
  {\path{doi:10.1103/PhysRevLett.127.131104}}.

\bibitem{Boskovic:2024fga}
Mateja Bo\v{s}kovi\'c, Matthias Koschnitzke, and Rafael~A. Porto.
\newblock {Signatures of Ultralight Bosons in the Orbital Eccentricity of
  Binary Black Holes}.
\newblock {\em Phys. Rev. Lett.}, 133(12):121401, 2024.
\newblock \href {http://arxiv.org/abs/2403.02415} {\path{arXiv:2403.02415}},
  \href {https://doi.org/10.1103/PhysRevLett.133.121401}
  {\path{doi:10.1103/PhysRevLett.133.121401}}.

\bibitem{Brito:2025ojt}
Richard Brito.
\newblock {Black holes as laboratories: searching for ultralight fields}.
\newblock {\em Gen. Rel. Grav.}, 57(2):42, 2025.
\newblock \href {https://doi.org/10.1007/s10714-025-03376-3}
  {\path{doi:10.1007/s10714-025-03376-3}}.

\bibitem{Doneva:2022ewd}
Daniela~D. Doneva, Fethi~M. Ramazano\u{g}lu, Hector~O. Silva, Thomas~P.
  Sotiriou, and Stoytcho~S. Yazadjiev.
\newblock {Spontaneous scalarization}.
\newblock {\em Rev. Mod. Phys.}, 96(1):015004, 2024.
\newblock \href {http://arxiv.org/abs/2211.01766} {\path{arXiv:2211.01766}},
  \href {https://doi.org/10.1103/RevModPhys.96.015004}
  {\path{doi:10.1103/RevModPhys.96.015004}}.

\bibitem{Damour:1993hw}
Thibault Damour and Gilles Esposito-Farese.
\newblock {Nonperturbative strong field effects in tensor - scalar theories of
  gravitation}.
\newblock {\em Phys. Rev. Lett.}, 70:2220--2223, 1993.
\newblock \href {https://doi.org/10.1103/PhysRevLett.70.2220}
  {\path{doi:10.1103/PhysRevLett.70.2220}}.

\bibitem{Sotiriou:2013qea}
Thomas~P. Sotiriou and Shuang-Yong Zhou.
\newblock {Black hole hair in generalized scalar-tensor gravity}.
\newblock {\em Phys. Rev. Lett.}, 112:251102, 2014.
\newblock \href {http://arxiv.org/abs/1312.3622} {\path{arXiv:1312.3622}},
  \href {https://doi.org/10.1103/PhysRevLett.112.251102}
  {\path{doi:10.1103/PhysRevLett.112.251102}}.

\bibitem{Antoniou:2017hxj}
G.~Antoniou, A.~Bakopoulos, and P.~Kanti.
\newblock {Black-Hole Solutions with Scalar Hair in
  Einstein-Scalar-Gauss-Bonnet Theories}.
\newblock {\em Phys. Rev. D}, 97(8):084037, 2018.
\newblock \href {http://arxiv.org/abs/1711.07431} {\path{arXiv:1711.07431}},
  \href {https://doi.org/10.1103/PhysRevD.97.084037}
  {\path{doi:10.1103/PhysRevD.97.084037}}.

\bibitem{Silva:2017uqg}
Hector~O. Silva, Jeremy Sakstein, Leonardo Gualtieri, Thomas~P. Sotiriou, and
  Emanuele Berti.
\newblock {Spontaneous scalarization of black holes and compact stars from a
  Gauss-Bonnet coupling}.
\newblock {\em Phys. Rev. Lett.}, 120(13):131104, 2018.
\newblock \href {http://arxiv.org/abs/1711.02080} {\path{arXiv:1711.02080}},
  \href {https://doi.org/10.1103/PhysRevLett.120.131104}
  {\path{doi:10.1103/PhysRevLett.120.131104}}.

\bibitem{Bekenstein:1998nt}
Jacob~D. Bekenstein and Marcelo Schiffer.
\newblock {The Many faces of superradiance}.
\newblock {\em Phys. Rev. D}, 58:064014, 1998.
\newblock \href {http://arxiv.org/abs/gr-qc/9803033}
  {\path{arXiv:gr-qc/9803033}}, \href
  {https://doi.org/10.1103/PhysRevD.58.064014}
  {\path{doi:10.1103/PhysRevD.58.064014}}.

\bibitem{Brito:2015oca}
Richard Brito, Vitor Cardoso, and Paolo Pani.
\newblock {\em {Superradiance}: {New Frontiers in Black Hole Physics}}, volume
  906.
\newblock Springer, 2015.
\newblock \href {http://arxiv.org/abs/1501.06570} {\path{arXiv:1501.06570}},
  \href {https://doi.org/10.1007/978-3-319-19000-6}
  {\path{doi:10.1007/978-3-319-19000-6}}.

\bibitem{Press:1972zz}
William~H. Press and Saul~A. Teukolsky.
\newblock {Floating Orbits, Superradiant Scattering and the Black-hole Bomb}.
\newblock {\em Nature}, 238:211--212, 1972.
\newblock \href {https://doi.org/10.1038/238211a0}
  {\path{doi:10.1038/238211a0}}.

\bibitem{Herdeiro:2013pia}
Carlos A.~R. Herdeiro, Juan~Carlos Degollado, and Helgi~Freyr R\'unarsson.
\newblock {Rapid growth of superradiant instabilities for charged black holes
  in a cavity}.
\newblock {\em Phys. Rev. D}, 88:063003, 2013.
\newblock \href {http://arxiv.org/abs/1305.5513} {\path{arXiv:1305.5513}},
  \href {https://doi.org/10.1103/PhysRevD.88.063003}
  {\path{doi:10.1103/PhysRevD.88.063003}}.

\bibitem{Cardoso:2004hs}
Vitor Cardoso and Oscar J.~C. Dias.
\newblock {Small Kerr-anti-de Sitter black holes are unstable}.
\newblock {\em Phys. Rev. D}, 70:084011, 2004.
\newblock \href {http://arxiv.org/abs/hep-th/0405006}
  {\path{arXiv:hep-th/0405006}}, \href
  {https://doi.org/10.1103/PhysRevD.70.084011}
  {\path{doi:10.1103/PhysRevD.70.084011}}.

\bibitem{Cardoso:2004nk}
Vitor Cardoso, Oscar J.~C. Dias, Jose P.~S. Lemos, and Shijun Yoshida.
\newblock {The Black hole bomb and superradiant instabilities}.
\newblock {\em Phys. Rev. D}, 70:044039, 2004.
\newblock [Erratum: Phys.Rev.D 70, 049903 (2004)].
\newblock \href {http://arxiv.org/abs/hep-th/0404096}
  {\path{arXiv:hep-th/0404096}}, \href
  {https://doi.org/10.1103/PhysRevD.70.049903}
  {\path{doi:10.1103/PhysRevD.70.049903}}.

\bibitem{Hui:2022sri}
Lam Hui, Y.~T.~Albert Law, Luca Santoni, Guanhao Sun, Giovanni~Maria Tomaselli,
  and Enrico Trincherini.
\newblock {Black hole superradiance with dark matter accretion}.
\newblock {\em Phys. Rev. D}, 107(10):104018, 2023.
\newblock \href {http://arxiv.org/abs/2208.06408} {\path{arXiv:2208.06408}},
  \href {https://doi.org/10.1103/PhysRevD.107.104018}
  {\path{doi:10.1103/PhysRevD.107.104018}}.

\bibitem{Guo:2025dkx}
Yin-Da Guo, Shou-Shan Bao, Tianjun Li, and Hong Zhang.
\newblock {The effect of accretion on scalar superradiant instability}.
\newblock 1 2025.
\newblock \href {http://arxiv.org/abs/2501.09280} {\path{arXiv:2501.09280}}.

\bibitem{Cardoso:2013opa}
Vitor Cardoso, Isabella~P. Carucci, Paolo Pani, and Thomas~P. Sotiriou.
\newblock {Matter around Kerr black holes in scalar-tensor theories:
  scalarization and superradiant instability}.
\newblock {\em Phys. Rev. D}, 88:044056, 2013.
\newblock \href {http://arxiv.org/abs/1305.6936} {\path{arXiv:1305.6936}},
  \href {https://doi.org/10.1103/PhysRevD.88.044056}
  {\path{doi:10.1103/PhysRevD.88.044056}}.

\bibitem{Silva:2020omi}
Hector~O. Silva, Helvi Witek, Matthew Elley, and Nicol\'as Yunes.
\newblock {Dynamical Descalarization in Binary Black Hole Mergers}.
\newblock {\em Phys. Rev. Lett.}, 127(3):031101, 2021.
\newblock \href {http://arxiv.org/abs/2012.10436} {\path{arXiv:2012.10436}},
  \href {https://doi.org/10.1103/PhysRevLett.127.031101}
  {\path{doi:10.1103/PhysRevLett.127.031101}}.

\bibitem{Corelli:2021ikv}
Fabrizio Corelli, Taishi Ikeda, and Paolo Pani.
\newblock {Challenging cosmic censorship in Einstein-Maxwell-scalar theory with
  numerically simulated gedanken experiments}.
\newblock {\em Phys. Rev. D}, 104(8):084069, 2021.
\newblock \href {http://arxiv.org/abs/2108.08328} {\path{arXiv:2108.08328}},
  \href {https://doi.org/10.1103/PhysRevD.104.084069}
  {\path{doi:10.1103/PhysRevD.104.084069}}.

\bibitem{Liu:2022eri}
Yunqi Liu, Cheng-Yong Zhang, Wei-Liang Qian, Kai Lin, and Bin Wang.
\newblock {Dynamic generation or removal of a scalar hair}.
\newblock {\em JHEP}, 01:074, 2023.
\newblock \href {http://arxiv.org/abs/2206.05012} {\path{arXiv:2206.05012}},
  \href {https://doi.org/10.1007/JHEP01(2023)074}
  {\path{doi:10.1007/JHEP01(2023)074}}.

\bibitem{Zhang:2022cmu}
Cheng-Yong Zhang, Qian Chen, Yunqi Liu, Wen-Kun Luo, Yu~Tian, and Bin Wang.
\newblock {Dynamical transitions in scalarization and descalarization through
  black hole accretion}.
\newblock {\em Phys. Rev. D}, 106(6):L061501, 2022.
\newblock \href {http://arxiv.org/abs/2204.09260} {\path{arXiv:2204.09260}},
  \href {https://doi.org/10.1103/PhysRevD.106.L061501}
  {\path{doi:10.1103/PhysRevD.106.L061501}}.

\bibitem{Stefanov:2007eq}
Ivan~Zh. Stefanov, Stoytcho~S. Yazadjiev, and Michail~D. Todorov.
\newblock {Phases of 4D scalar-tensor black holes coupled to Born-Infeld
  nonlinear electrodynamics}.
\newblock {\em Mod. Phys. Lett. A}, 23:2915--2931, 2008.
\newblock \href {http://arxiv.org/abs/0708.4141} {\path{arXiv:0708.4141}},
  \href {https://doi.org/10.1142/S0217732308028351}
  {\path{doi:10.1142/S0217732308028351}}.

\bibitem{Herdeiro:2018wub}
Carlos~A.R. Herdeiro, Eugen Radu, Nicolas Sanchis-Gual, and Jos\'e~A. Font.
\newblock {Spontaneous Scalarization of Charged Black Holes}.
\newblock {\em Phys. Rev. Lett.}, 121(10):101102, 2018.
\newblock \href {http://arxiv.org/abs/1806.05190} {\path{arXiv:1806.05190}},
  \href {https://doi.org/10.1103/PhysRevLett.121.101102}
  {\path{doi:10.1103/PhysRevLett.121.101102}}.

\bibitem{DiMenza:2014vpa}
Laurent Di~Menza and Jean-Philippe Nicolas.
\newblock {Superradiance on the Reissner\textendash{}Nordstr\o{}m metric}.
\newblock {\em Class. Quant. Grav.}, 32(14):145013, 2015.
\newblock \href {http://arxiv.org/abs/1411.3988} {\path{arXiv:1411.3988}},
  \href {https://doi.org/10.1088/0264-9381/32/14/145013}
  {\path{doi:10.1088/0264-9381/32/14/145013}}.

\bibitem{Benone:2015bst}
Carolina~L. Benone and Lu\'\i{}s C.~B. Crispino.
\newblock {Superradiance in static black hole spacetimes}.
\newblock {\em Phys. Rev. D}, 93(2):024028, 2016.
\newblock \href {http://arxiv.org/abs/1511.02634} {\path{arXiv:1511.02634}},
  \href {https://doi.org/10.1103/PhysRevD.93.024028}
  {\path{doi:10.1103/PhysRevD.93.024028}}.

\bibitem{Baake:2016oku}
Olaf Baake and Oliver Rinne.
\newblock {Superradiance of a charged scalar field coupled to the
  Einstein-Maxwell equations}.
\newblock {\em Phys. Rev. D}, 94(12):124016, 2016.
\newblock \href {http://arxiv.org/abs/1610.08352} {\path{arXiv:1610.08352}},
  \href {https://doi.org/10.1103/PhysRevD.94.124016}
  {\path{doi:10.1103/PhysRevD.94.124016}}.

\bibitem{Garcia-Saenz:2024beb}
Sebastian Garcia-Saenz, Guangzhou Guo, Peng Wang, and Xinmiao Wang.
\newblock {Black hole accretion of scalar clouds with spontaneous symmetry
  breaking}.
\newblock {\em Phys. Rev. D}, 110(12):124045, 2024.
\newblock \href {http://arxiv.org/abs/2409.13184} {\path{arXiv:2409.13184}},
  \href {https://doi.org/10.1103/PhysRevD.110.124045}
  {\path{doi:10.1103/PhysRevD.110.124045}}.

\bibitem{Essig:2013lka}
Rouven Essig et~al.
\newblock {Working Group Report: New Light Weakly Coupled Particles}.
\newblock In {\em {Snowmass 2013}: {Snowmass on the Mississippi}}, 10 2013.
\newblock \href {http://arxiv.org/abs/1311.0029} {\path{arXiv:1311.0029}}.

\bibitem{Hartnoll:2009sz}
Sean~A. Hartnoll.
\newblock {Lectures on holographic methods for condensed matter physics}.
\newblock {\em Class. Quant. Grav.}, 26:224002, 2009.
\newblock \href {http://arxiv.org/abs/0903.3246} {\path{arXiv:0903.3246}},
  \href {https://doi.org/10.1088/0264-9381/26/22/224002}
  {\path{doi:10.1088/0264-9381/26/22/224002}}.

\bibitem{Herzog:2009xv}
Christopher~P. Herzog.
\newblock {Lectures on Holographic Superfluidity and Superconductivity}.
\newblock {\em J. Phys. A}, 42:343001, 2009.
\newblock \href {http://arxiv.org/abs/0904.1975} {\path{arXiv:0904.1975}},
  \href {https://doi.org/10.1088/1751-8113/42/34/343001}
  {\path{doi:10.1088/1751-8113/42/34/343001}}.

\bibitem{Horowitz:2010gk}
Gary~T. Horowitz.
\newblock {Introduction to Holographic Superconductors}.
\newblock {\em Lect. Notes Phys.}, 828:313--347, 2011.
\newblock \href {http://arxiv.org/abs/1002.1722} {\path{arXiv:1002.1722}},
  \href {https://doi.org/10.1007/978-3-642-04864-7_10}
  {\path{doi:10.1007/978-3-642-04864-7_10}}.

\bibitem{Hod:2013nn}
Shahar Hod.
\newblock {No-bomb theorem for charged Reissner-Nordstroem black holes}.
\newblock {\em Phys. Lett. B}, 718:1489--1492, 2013.
\newblock \href {https://doi.org/10.1016/j.physletb.2012.12.013}
  {\path{doi:10.1016/j.physletb.2012.12.013}}.

\bibitem{Hod:2015hza}
Shahar Hod.
\newblock {Stability of highly-charged Reissner-Nordstr\"om black holes to
  charged scalar perturbations}.
\newblock {\em Phys. Rev. D}, 91(4):044047, 2015.
\newblock \href {http://arxiv.org/abs/1504.00009} {\path{arXiv:1504.00009}},
  \href {https://doi.org/10.1103/PhysRevD.91.044047}
  {\path{doi:10.1103/PhysRevD.91.044047}}.

\bibitem{Fernandes:2019rez}
Pedro G.~S. Fernandes, Carlos A.~R. Herdeiro, Alexandre~M. Pombo, Eugen Radu,
  and Nicolas Sanchis-Gual.
\newblock {Spontaneous Scalarisation of Charged Black Holes: Coupling
  Dependence and Dynamical Features}.
\newblock {\em Class. Quant. Grav.}, 36(13):134002, 2019.
\newblock [Erratum: Class.Quant.Grav. 37, 049501 (2020)].
\newblock \href {http://arxiv.org/abs/1902.05079} {\path{arXiv:1902.05079}},
  \href {https://doi.org/10.1088/1361-6382/ab23a1}
  {\path{doi:10.1088/1361-6382/ab23a1}}.

\bibitem{Minamitsuji:2021vdb}
Masato Minamitsuji and Shinji Tsujikawa.
\newblock {Spontaneous scalarization of charged stars}.
\newblock {\em Phys. Lett. B}, 820:136509, 2021.
\newblock \href {http://arxiv.org/abs/2105.14661} {\path{arXiv:2105.14661}},
  \href {https://doi.org/10.1016/j.physletb.2021.136509}
  {\path{doi:10.1016/j.physletb.2021.136509}}.

\bibitem{Guo:2023mda}
Guangzhou Guo, Peng Wang, Houwen Wu, and Haitang Yang.
\newblock {Scalarized Kerr-Newman black holes}.
\newblock {\em JHEP}, 10:076, 2023.
\newblock \href {http://arxiv.org/abs/2307.12210} {\path{arXiv:2307.12210}},
  \href {https://doi.org/10.1007/JHEP10(2023)076}
  {\path{doi:10.1007/JHEP10(2023)076}}.

\bibitem{Melis:2024kfr}
Marco Melis, Fabrizio Corelli, Robin Croft, and Paolo Pani.
\newblock {Black hole spectroscopy and nonlinear echoes in
  Einstein-Maxwell-scalar theory}.
\newblock 12 2024.
\newblock \href {http://arxiv.org/abs/2412.14259} {\path{arXiv:2412.14259}}.

\bibitem{Latosh:2023cxm}
Boris Latosh and Miok Park.
\newblock {Hairy black holes by spontaneous symmetry breaking}.
\newblock {\em Phys. Rev. D}, 110(2):024012, 2024.
\newblock \href {http://arxiv.org/abs/2305.19814} {\path{arXiv:2305.19814}},
  \href {https://doi.org/10.1103/PhysRevD.110.024012}
  {\path{doi:10.1103/PhysRevD.110.024012}}.

\bibitem{Hyun:2024sfv}
Young-Hwan Hyun, Boris Latosh, and Miok Park.
\newblock {Scalar field perturbation of hairy black holes in EsGB theory}.
\newblock {\em JHEP}, 08:163, 2024.
\newblock \href {http://arxiv.org/abs/2405.08769} {\path{arXiv:2405.08769}},
  \href {https://doi.org/10.1007/JHEP08(2024)163}
  {\path{doi:10.1007/JHEP08(2024)163}}.

\bibitem{Baumgarte:2010ndz}
Thomas~W. Baumgarte and Stuart~L. Shapiro.
\newblock {\em {Numerical Relativity: Solving Einstein's Equations on the
  Computer}}.
\newblock Cambridge University Press, 2010.
\newblock \href {https://doi.org/10.1017/CBO9781139193344}
  {\path{doi:10.1017/CBO9781139193344}}.

\bibitem{Frauendiener:2011zz}
Jorg Frauendiener.
\newblock {Miguel Alcubierre: Introduction to 3 + 1 numerical relativity}.
\newblock {\em Gen. Rel. Grav.}, 43:2931--2933, 2011.
\newblock \href {https://doi.org/10.1007/s10714-011-1195-5}
  {\path{doi:10.1007/s10714-011-1195-5}}.

\bibitem{Torres:2014fga}
Jose~M. Torres and Miguel Alcubierre.
\newblock {Gravitational collapse of charged scalar fields}.
\newblock {\em Gen. Rel. Grav.}, 46:1773, 2014.
\newblock \href {http://arxiv.org/abs/1407.7885} {\path{arXiv:1407.7885}},
  \href {https://doi.org/10.1007/s10714-014-1773-4}
  {\path{doi:10.1007/s10714-014-1773-4}}.

\bibitem{Sanchis-Gual:2016tcm}
Nicolas Sanchis-Gual, Juan~Carlos Degollado, Carlos Herdeiro, Jos\'e~A. Font,
  and Pedro~J. Montero.
\newblock {Dynamical formation of a Reissner-Nordstr\"om black hole with scalar
  hair in a cavity}.
\newblock {\em Phys. Rev. D}, 94(4):044061, 2016.
\newblock \href {http://arxiv.org/abs/1607.06304} {\path{arXiv:1607.06304}},
  \href {https://doi.org/10.1103/PhysRevD.94.044061}
  {\path{doi:10.1103/PhysRevD.94.044061}}.

\bibitem{Hirschmann:2017psw}
Eric~W. Hirschmann, Luis Lehner, Steven~L. Liebling, and Carlos Palenzuela.
\newblock {Black Hole Dynamics in Einstein-Maxwell-Dilaton Theory}.
\newblock {\em Phys. Rev. D}, 97(6):064032, 2018.
\newblock \href {http://arxiv.org/abs/1706.09875} {\path{arXiv:1706.09875}},
  \href {https://doi.org/10.1103/PhysRevD.97.064032}
  {\path{doi:10.1103/PhysRevD.97.064032}}.

\bibitem{Garcia-Saenz:2025dsr}
Sebastian Garcia-Saenz, Guangzhou Guo, Peng Wang, and Xinmiao Wang.
\newblock {Stable long-term evolution in numerical relativity}.
\newblock 1 2025.
\newblock \href {http://arxiv.org/abs/2501.01055} {\path{arXiv:2501.01055}}.

\bibitem{Misner:1974qy}
Charles~W. Misner, K.~S. Thorne, and J.~A. Wheeler.
\newblock {\em {Gravitation}}.
\newblock W. H. Freeman, San Francisco, 1973.

\bibitem{Poisson:2009pwt}
Eric Poisson.
\newblock {\em {A Relativist's Toolkit: The Mathematics of Black-Hole
  Mechanics}}.
\newblock Cambridge University Press, 12 2009.
\newblock \href {https://doi.org/10.1017/CBO9780511606601}
  {\path{doi:10.1017/CBO9780511606601}}.

\bibitem{Clough:2021qlv}
Katy Clough.
\newblock {Continuity equations for general matter: applications in numerical
  relativity}.
\newblock {\em Class. Quant. Grav.}, 38(16):167001, 2021.
\newblock \href {http://arxiv.org/abs/2104.13420} {\path{arXiv:2104.13420}},
  \href {https://doi.org/10.1088/1361-6382/ac10ee}
  {\path{doi:10.1088/1361-6382/ac10ee}}.

\bibitem{Croft:2022gks}
Robin Croft.
\newblock {Local continuity of angular momentum and noether charge for matter
  in general relativity}.
\newblock {\em Class. Quant. Grav.}, 40(10):105007, 2023.
\newblock \href {http://arxiv.org/abs/2203.13845} {\path{arXiv:2203.13845}},
  \href {https://doi.org/10.1088/1361-6382/accc6a}
  {\path{doi:10.1088/1361-6382/accc6a}}.

\bibitem{Guo:2023ivz}
Guangzhou Guo, Peng Wang, Houwen Wu, and Haitang Yang.
\newblock {Superradiance instabilities of charged black holes in
  Einstein-Maxwell-scalar theory}.
\newblock {\em JHEP}, 07:070, 2023.
\newblock \href {http://arxiv.org/abs/2301.06483} {\path{arXiv:2301.06483}},
  \href {https://doi.org/10.1007/JHEP07(2023)070}
  {\path{doi:10.1007/JHEP07(2023)070}}.

\bibitem{Hod:2012wmy}
Shahar Hod.
\newblock {Stability of the extremal Reissner-Nordstroem black hole to charged
  scalar perturbations}.
\newblock {\em Phys. Lett. B}, 713:505--508, 2012.
\newblock \href {http://arxiv.org/abs/1304.6474} {\path{arXiv:1304.6474}},
  \href {https://doi.org/10.1016/j.physletb.2012.06.043}
  {\path{doi:10.1016/j.physletb.2012.06.043}}.

\bibitem{Zhu:2014sya}
Zhiying Zhu, Shao-Jun Zhang, C.~E. Pellicer, Bin Wang, and Elcio Abdalla.
\newblock {Stability of Reissner-Nordstr\"om black hole in de Sitter background
  under charged scalar perturbation}.
\newblock {\em Phys. Rev. D}, 90(4):044042, 2014.
\newblock [Addendum: Phys.Rev.D 90, 049904 (2014)].
\newblock \href {http://arxiv.org/abs/1405.4931} {\path{arXiv:1405.4931}},
  \href {https://doi.org/10.1103/PhysRevD.90.044042}
  {\path{doi:10.1103/PhysRevD.90.044042}}.

\bibitem{Huang:2015cha}
Yang Huang and Dao-Jun Liu.
\newblock {Charged scalar perturbations around a regular magnetic black hole}.
\newblock {\em Phys. Rev. D}, 93(10):104011, 2016.
\newblock \href {http://arxiv.org/abs/1509.09017} {\path{arXiv:1509.09017}},
  \href {https://doi.org/10.1103/PhysRevD.93.104011}
  {\path{doi:10.1103/PhysRevD.93.104011}}.

\bibitem{Sanchis-Gual:2015lje}
Nicolas Sanchis-Gual, Juan~Carlos Degollado, Pedro~J. Montero, Jos\'e~A. Font,
  and Carlos Herdeiro.
\newblock {Explosion and Final State of an Unstable Reissner-Nordstr\"om Black
  Hole}.
\newblock {\em Phys. Rev. Lett.}, 116(14):141101, 2016.
\newblock \href {http://arxiv.org/abs/1512.05358} {\path{arXiv:1512.05358}},
  \href {https://doi.org/10.1103/PhysRevLett.116.141101}
  {\path{doi:10.1103/PhysRevLett.116.141101}}.

\bibitem{BHathome}
Zachariah~B. Etienne and Ian~Ruchlin et~al.
\newblock {BlackHoles@Home, 2022, To find out more, visit
  https://blackholesathome.net/}.

\bibitem{Campanelli:2005dd}
Manuela Campanelli, C.~O. Lousto, P.~Marronetti, and Y.~Zlochower.
\newblock {Accurate evolutions of orbiting black-hole binaries without
  excision}.
\newblock {\em Phys. Rev. Lett.}, 96:111101, 2006.
\newblock \href {http://arxiv.org/abs/gr-qc/0511048}
  {\path{arXiv:gr-qc/0511048}}, \href
  {https://doi.org/10.1103/PhysRevLett.96.111101}
  {\path{doi:10.1103/PhysRevLett.96.111101}}.

\bibitem{Baker:2005vv}
John~G. Baker, Joan Centrella, Dae-Il Choi, Michael Koppitz, and James van
  Meter.
\newblock {Gravitational wave extraction from an inspiraling configuration of
  merging black holes}.
\newblock {\em Phys. Rev. Lett.}, 96:111102, 2006.
\newblock \href {http://arxiv.org/abs/gr-qc/0511103}
  {\path{arXiv:gr-qc/0511103}}, \href
  {https://doi.org/10.1103/PhysRevLett.96.111102}
  {\path{doi:10.1103/PhysRevLett.96.111102}}.

\bibitem{Brown:2009ki}
J.~David Brown.
\newblock {Probing the puncture for black hole simulations}.
\newblock {\em Phys. Rev. D}, 80:084042, 2009.
\newblock \href {http://arxiv.org/abs/0908.3814} {\path{arXiv:0908.3814}},
  \href {https://doi.org/10.1103/PhysRevD.80.084042}
  {\path{doi:10.1103/PhysRevD.80.084042}}.

\bibitem{Herdeiro:2015waa}
Carlos~A.R. Herdeiro and Eugen Radu.
\newblock {Asymptotically flat black holes with scalar hair: a review}.
\newblock {\em Int. J. Mod. Phys. D}, 24(09):1542014, 2015.
\newblock \href {http://arxiv.org/abs/1504.08209} {\path{arXiv:1504.08209}},
  \href {https://doi.org/10.1142/S0218271815420146}
  {\path{doi:10.1142/S0218271815420146}}.

\bibitem{Mayo:1996mv}
Avraham~E. Mayo and Jacob~D. Bekenstein.
\newblock {No hair for spherical black holes: Charged and nonminimally coupled
  scalar field with selfinteraction}.
\newblock {\em Phys. Rev. D}, 54:5059--5069, 1996.
\newblock \href {http://arxiv.org/abs/gr-qc/9602057}
  {\path{arXiv:gr-qc/9602057}}, \href
  {https://doi.org/10.1103/PhysRevD.54.5059}
  {\path{doi:10.1103/PhysRevD.54.5059}}.

\bibitem{Hong:2020miv}
Jeong-Pyong Hong, Motoo Suzuki, and Masaki Yamada.
\newblock {Spherically Symmetric Scalar Hair for Charged Black Holes}.
\newblock {\em Phys. Rev. Lett.}, 125(11):111104, 2020.
\newblock \href {http://arxiv.org/abs/2004.03148} {\path{arXiv:2004.03148}},
  \href {https://doi.org/10.1103/PhysRevLett.125.111104}
  {\path{doi:10.1103/PhysRevLett.125.111104}}.

\bibitem{Herdeiro:2020xmb}
Carlos A.~R. Herdeiro and Eugen Radu.
\newblock {Spherical electro-vacuum black holes with resonant, scalar
  $Q$-hair}.
\newblock {\em Eur. Phys. J. C}, 80(5):390, 2020.
\newblock \href {http://arxiv.org/abs/2004.00336} {\path{arXiv:2004.00336}},
  \href {https://doi.org/10.1140/epjc/s10052-020-7976-9}
  {\path{doi:10.1140/epjc/s10052-020-7976-9}}.

\bibitem{Hartnoll:2020fhc}
Sean~A. Hartnoll, Gary~T. Horowitz, Jorrit Kruthoff, and Jorge~E. Santos.
\newblock {Diving into a holographic superconductor}.
\newblock {\em SciPost Phys.}, 10(1):009, 2021.
\newblock \href {http://arxiv.org/abs/2008.12786} {\path{arXiv:2008.12786}},
  \href {https://doi.org/10.21468/SciPostPhys.10.1.009}
  {\path{doi:10.21468/SciPostPhys.10.1.009}}.

\bibitem{Cai:2020wrp}
Rong-Gen Cai, Li~Li, and Run-Qiu Yang.
\newblock {No Inner-Horizon Theorem for Black Holes with Charged Scalar Hairs}.
\newblock {\em JHEP}, 03:263, 2021.
\newblock \href {http://arxiv.org/abs/2009.05520} {\path{arXiv:2009.05520}},
  \href {https://doi.org/10.1007/JHEP03(2021)263}
  {\path{doi:10.1007/JHEP03(2021)263}}.

\bibitem{An:2023bpb}
Yu-Ping An and Li~Li.
\newblock {Static de-Sitter black holes abhor charged scalar hair}.
\newblock {\em Eur. Phys. J. C}, 83(7):569, 2023.
\newblock \href {http://arxiv.org/abs/2301.06312} {\path{arXiv:2301.06312}},
  \href {https://doi.org/10.1140/epjc/s10052-023-11758-7}
  {\path{doi:10.1140/epjc/s10052-023-11758-7}}.

\bibitem{Herdeiro:2024yqa}
Carlos Herdeiro, Eugen Radu, and Yakov Shnir.
\newblock {Reissner-Nordstr\"om dyonic black holes with gauged scalar hair}.
\newblock {\em Phys. Lett. B}, 856:138912, 2024.
\newblock \href {http://arxiv.org/abs/2406.10643} {\path{arXiv:2406.10643}},
  \href {https://doi.org/10.1016/j.physletb.2024.138912}
  {\path{doi:10.1016/j.physletb.2024.138912}}.

\end{thebibliography}

\end{document}